\documentclass[preprint,aps,12pt,showpacs,nofootinbib,tightenlines]{revtex4}
\usepackage{amsmath}
\usepackage{amssymb}
\usepackage{epsfig}
\usepackage{graphicx}
\begin{document}

\newcommand{\beq}{\begin{eqnarray}}
\newcommand{\eeq}{\end{eqnarray}}
\newcommand{\non}{\nonumber\\ }

\newcommand{\acp}{{\cal A}_{CP}}
\newcommand{\etap}{\eta^{(\prime)} }
\newcommand{\pb}{\phi_B}
\newcommand{\pp}{\phi_{\pi}}
\newcommand{\pe}{\phi_{\eta}}
\newcommand{\pepr}{\phi_{\etap}}
\newcommand{\ppp}{\phi_{\pi}^P}
\newcommand{\pep}{\phi_{\eta}^P}
\newcommand{\peprp}{\phi_{\etap}^P}
\newcommand{\ppt}{\phi_{\pi}^t}
\newcommand{\pet}{\phi_{\eta}^t}
\newcommand{\peprt}{\phi_{\etap}^t}
\newcommand{\fb}{f_B }
\newcommand{\fpi}{f_{\pi} }
\newcommand{\feta}{f_{\eta} }
\newcommand{\fetap}{f_{\etap} }
\newcommand{\rpi}{r_{\pi} }
\newcommand{\re}{r_{\eta} }
\newcommand{\rep}{r_{\etap} }
\newcommand{\mb}{m_B }
\newcommand{\mop}{m_{0\pi} }
\newcommand{\moe}{m_{0\eta} }
\newcommand{\moep}{m_{0\etap} }

\newcommand{\psl}{ p \hspace{-1.8truemm}/ }
\newcommand{\nsl}{ n \hspace{-2.2truemm}/ }
\newcommand{\vsl}{ v \hspace{-2.2truemm}/ }
\newcommand{\epsl}{\epsilon \hspace{-1.8truemm}/\,  }

\def \epjc{ Eur. Phys. J. C }
\def \jpg{  J. Phys. G }
\def \npb{  Nucl. Phys. B }
\def \plb{  Phys. Lett. B }
\def \pr{  Phys. Rep. }
\def \prd{  Phys. Rev. D }
\def \prl{  Phys. Rev. Lett.  }
\def \zpc{  Z. Phys. C  }
\def \jhep{ J. High Energy Phys.  }

\title{Branching Ratio and CP Asymmetry of $B \to
\pi \eta^{(\prime)}$ Decays in the Perturbative QCD Approach}
\author{Huisheng Wang, Xin Liu}
\author{Zhenjun Xiao (corresponding author)}\email{xiaozhenjun@njnu.edu.cn}
\author{Libo Guo}\email{guolb@email.njnu.edu.cn}
\affiliation{Department of Physics and Institute of Theoretical Physics,
Nanjing Normal University, Nanjing, Jiangsu 210097, P.R.China}
\author{ Cai-Dian L\"u}\email{lucd@ihep.ac.cn}
\affiliation{ CCAST(World Laboratory), P.O. Box 8730, Beijing
100080, P.R. China; }
\affiliation{Institute of High Energy Physics,
CAS, P.O.Box 918(4) Beijing 100049,   China
\footnote{Mailing address.}} 
\date{\today}
\begin{abstract}
We calculate the branching ratios and CP-violating asymmetries for
$B^0 \to \pi^0 \eta^{(\prime)}$ and $B^+\to \pi^+ \eta^{(\prime)}$
decays in the perturbative QCD (pQCD) factorization approach here.
We not only calculate the usual factorizable contributions, but
also evaluate the non-factorizable and annihilation type
contributions. Besides the current-current operators, the
contributions from the QCD and electroweak penguin operators are
also taken into account. The pQCD results for the CP-averaged
branching ratios are $Br(B^+ \to \pi^+ \eta) \approx 4.1 \times
10^{-6}$, $Br(B^+ \to \pi^+ \eta^\prime) \approx 2.4 \times
10^{-6}$,
 and $Br(B^0 \to \pi^0 \eta^{(\prime)}) \approx 0.2 \times 10^{-6}$,
which agree very well with the measured values or currently
available experimental upper limits. We also predict large
CP-violating  asymmetries in these decays: $A_{CP}^{dir}(\pi^\pm
\etap)\sim A_{CP}^{dir}(\pi^0 \etap)\sim -0.35  $, and
$A_{CP}^{mix}(\pi^0 \etap)\sim 0.67$, but with large errors.
The pQCD prediction for $A_{CP}^{dir}(\pi^\pm \eta)$ ($A_{CP}^{dir}(\pi^\pm
\eta^\prime))$ has the same (opposite) sign with the primary measured values.
Further improvements in both theory and  experiments are needed
to clarify this discrepancy.
\end{abstract}

\vspace{1cm}


\maketitle

\section{Introduction}

The experimental measurements and theoretical studies of the two
body charmless hadronic B meson decays play an important role in
the precision test of the standard mode (SM) and in searching for
the new physics beyond the SM \cite{cpv}. For these charmless B
meson decays, the dominant theoretical error comes from the large
uncertainty in evaluating the hadronic matrix elements $\langle
M_1 M_2|O_i|B\rangle$ where $M_1$ and $M_2$ are light final state
mesons. The QCD factorization (QCDF) approach \cite{bbns99} and
the perturbative QCD (pQCD) factorization approach
\cite{cl97,li2003,lb80} are the popular methods being used to
calculate the hadronic matrix element.  Many charmless B meson
decays, for example, have been calculated in the QCDF  approach
\cite{bbns99,du03,yy01,bn03b} and/or in the pQCD approach
\cite{luy01,kls01,li01,kklls04,li05,liu05}.

In this paper, we would like to calculate the branching ratios and
CP asymmetries for the four $B \to \pi \eta^{(\prime)}$ decays by
employing the low energy effective Hamiltonian \cite{buras96} and
the pQCD approach. Besides the usual factorizable contributions,
we here are able to evaluate the non-factorizable and the
annihilation contributions to these decays. Theoretically, the
four $B \to \pi \eta^{(\prime)}$ decays have been studied before
in  the naive or generalized factorization approach \cite{ali98}
as well as in the QCDF approach \cite{du03}. On the experimental
side, the CP-averaged branching ratios and CP-violating
asymmetries of $B \to \pi^+ \eta$ and $\pi^+ \eta^\prime$ decays
have been measured very recently \cite{babar,belle}, and the world
averages as given by the Heavy Flavor Averaging Group \cite{hfag}
are \beq Br(B^+ \to \pi^+ \eta )&=& \left ( 4.3 \pm
0.4\right )\times 10^{-6}\; ,\label{eq:exp1}\\
Br(B^+ \to \pi^+ \eta^\prime )&=& \left ( 2.53 ^{+0.59}_{-0.50}
 \right )\times 10^{-6} \; ,  \label{eq:exp2}\\
\acp(B^\pm \to \pi^\pm \eta )&=& -0.11 \pm 0.08\; ,\label{eq:acpexp1}\\
\acp(B^+ \to \pi^+ \eta^\prime )&=& 0.14 \pm 0.15 \; .  \label{eq:acpexp2}
\eeq
For $B \to \pi^0 \eta, \pi^0 \eta^\prime$ decays, only the
experimental upper limits for the branching ratios are available now \cite{hfag}
\beq
Br(B^0 \to \pi^0 \eta)< 2.5 \times 10^{-6}, \quad Br(B^0 \to \pi^0
\eta^\prime)< 3.7 \times 10^{-6}. \label{eq:ulimits}
\eeq

For $B \to \pi \eta^{(\prime)}$ decays, the $B$ meson is heavy,
setting at rest and decaying into two light mesons (i.e. $\pi$ and
$\eta^{(\prime)}$) with large momenta. Therefore the light final
state mesons are moving very fast in the rest frame of $B$ meson.
In this case, the short distance hard process dominates the decay
amplitude. We shall demonstrate that the soft final state
interaction is not important for such decays, since there is not
enough time for light mesons to exchange soft gluons. Therefore,
it makes the pQCD reliable in calculating the $B \to \pi
\eta^{(\prime)}$ decays. With the Sudakov resummation, we can
include the leading double logarithms for all loop diagrams, in
association with the soft contribution. Unlike the usual
factorization approach, the hard part of the pQCD approach
consists of six quarks rather than four. We thus call it six-quark
operators or six-quark effective theory.  Applying the six-quark
effective theory to B meson decays, we need meson wave functions
for the hadronization of quarks into mesons. All the collinear
dynamics are included in the meson wave functions.

This paper is organized as follows. In Sec.~\ref{sec:f-work}, we
give a brief review for the PQCD factorization approach. In
Sec.~\ref{sec:p-c}, we calculate analytically the related Feynman
diagrams and present the various decay amplitudes for the studied
decay modes. In Sec.~\ref{sec:n-d}, we show the numerical results
for the branching ratios and CP asymmetries of $B \to \pi
\eta^{(')}$ decays and compare them with the measured values or
the theoretical predictions in QCDF approach.
The summary and some discussions are included in the final
section.

\section{ Theoretical framework}\label{sec:f-work}

The three scale PQCD factorization approach has been developed and
applied in the non-leptonic $B$ meson decays
\cite{cl97,li2003,lb80,luy01,kls01,li01,kklls04} for some time. In
this approach, the decay amplitude is separated into soft, hard,
and harder dynamics characterized by different energy scales $(t,
m_b, M_W)$. It is conceptually written as the convolution, \beq
{\cal A}(B \to M_1 M_2)\sim \int\!\! d^4k_1 d^4k_2 d^4k_3\
\mathrm{Tr} \left [ C(t) \Phi_B(k_1) \Phi_{M_1}(k_2)
\Phi_{M_2}(k_3) H(k_1,k_2,k_3, t) \right ], \label{eq:con1} \eeq
where $k_i$'s are momenta of light quarks included in each mesons,
and $\mathrm{Tr}$ denotes the trace over Dirac and color indices.
$C(t)$ is the Wilson coefficient which results from the radiative
corrections at short distance. In the above convolution, $C(t)$
includes the harder dynamics at larger scale than $M_B$ scale and
describes the evolution of local $4$-Fermi operators from $m_W$
(the $W$ boson mass) down to $t\sim\mathcal{O}(\sqrt{\bar{\Lambda}
M_B})$ scale, where $\bar{\Lambda}\equiv M_B -m_b$. The function
$H(k_1,k_2,k_3,t)$ describes the four quark operator and the
spectator quark connected by
 a hard gluon whose $q^2$ is in the order
of $\bar{\Lambda} M_B$, and includes the
$\mathcal{O}(\sqrt{\bar{\Lambda} M_B})$ hard dynamics. Therefore,
this hard part $H$ can be perturbatively calculated. The function $\Phi_M$ is
the wave function which describes hadronization of the quark and
anti-quark to the meson $M$. While the function $H$ depends on the
processes considered, the wave function $\Phi_M$ is  independent of the specific
processes. Using the wave functions determined from other well measured
processes, one can make quantitative predictions here.

Since the b quark is rather heavy we consider the $B$ meson at
rest for simplicity. It is convenient to use light-cone coordinate
$(p^+, p^-, {\bf p}_T)$ to describe the meson's momenta, \beq
p^\pm = \frac{1}{\sqrt{2}} (p^0 \pm p^3), \quad and \quad {\bf
p}_T = (p^1, p^2). \eeq Using these coordinates the $B$ meson and
the two final state meson momenta can be written as \beq P_1 =
\frac{M_B}{\sqrt{2}} (1,1,{\bf 0}_T), \quad P_2 =
\frac{M_B}{\sqrt{2}}(1,0,{\bf 0}_T), \quad P_3 =
\frac{M_B}{\sqrt{2}} (0,1,{\bf 0}_T), \eeq respectively, here the
light meson masses have been neglected. Putting the light (anti-)
quark momenta in $B$, $\pi$ and $\eta$ mesons as $k_1$, $k_2$, and
$k_3$, respectively, we can choose \beq k_1 = (x_1 P_1^+,0,{\bf
k}_{1T}), \quad k_2 = (x_2 P_2^+,0,{\bf k}_{2T}), \quad k_3 = (0,
x_3 P_3^-,{\bf k}_{3T}). \eeq Then, the integration over $k_1^-$,
$k_2^-$, and $k_3^+$ in eq.(\ref{eq:con1}) will lead to \beq {\cal
A}(B \to \pi \etap) &\sim &\int\!\! d x_1 d x_2 d x_3 b_1 d b_1
b_2 d b_2 b_3 d b_3 \non && \cdot \mathrm{Tr} \left [ C(t)
\Phi_B(x_1,b_1) \Phi_\pi(x_2,b_2) \Phi_{\etap}(x_3, b_3) H(x_i,
b_i, t) S_t(x_i)\, e^{-S(t)} \right ], \label{eq:a2} \eeq where
$b_i$ is the conjugate space coordinate of $k_{iT}$, and $t$ is
the largest energy scale in function $H(x_i,b_i,t)$. The large
logarithms ($\ln m_W/t$) coming from QCD radiative corrections to
four quark operators are included in the Wilson coefficients
$C(t)$. The large double logarithms ($\ln^2 x_i$) on the
longitudinal direction are summed by the threshold resummation
\cite{li02}, and they lead to $S_t(x_i)$ which smears the
end-point singularities on $x_i$. The last term, $e^{-S(t)}$, is
the Sudakov form factor which suppresses the soft dynamics
effectively \cite{soft}. Thus it makes the perturbative
calculation of the hard part $H$ applicable at intermediate scale,
i.e., $M_B$ scale. We will calculate analytically the function
$H(x_i,b_i,t)$ for $B \to \pi \etap$ decays in the first order in
$\alpha_s$ expansion and give the convoluted amplitudes in next
section.

\subsection{ Wilson Coefficients}\label{ssec:w-c}

For $B \to \pi \etap$ decays, the related weak effective
Hamiltonian $H_{eff}$ can be written as \cite{buras96}
\beq
\label{eq:heff} {\cal H}_{eff} = \frac{G_{F}} {\sqrt{2}} \, \left[
V_{ub} V_{ud}^* \left (C_1(\mu) O_1^u(\mu) + C_2(\mu) O_2^u(\mu)
\right) - V_{tb} V_{td}^* \, \sum_{i=3}^{10} C_{i}(\mu) \,O_i(\mu)
\right] \; .
\eeq
We specify below the operators in ${\cal H}_{eff}$ for $b \to d$ transition:
\beq
\begin{array}{llllll}
O_1^{u} & = &  \bar d_\alpha\gamma^\mu L u_\beta\cdot \bar
u_\beta\gamma_\mu L b_\alpha\ , &O_2^{u} & = &\bar
d_\alpha\gamma^\mu L u_\alpha\cdot \bar
u_\beta\gamma_\mu L b_\beta\ , \\
O_3 & = & \bar d_\alpha\gamma^\mu L b_\alpha\cdot \sum_{q'}\bar
 q_\beta'\gamma_\mu L q_\beta'\ ,   &
O_4 & = & \bar d_\alpha\gamma^\mu L b_\beta\cdot \sum_{q'}\bar
q_\beta'\gamma_\mu L q_\alpha'\ , \\
O_5 & = & \bar d_\alpha\gamma^\mu L b_\alpha\cdot \sum_{q'}\bar
q_\beta'\gamma_\mu R q_\beta'\ ,   & O_6 & = & \bar
d_\alpha\gamma^\mu L b_\beta\cdot \sum_{q'}\bar
q_\beta'\gamma_\mu R q_\alpha'\ , \\
O_7 & = & \frac{3}{2}\bar d_\alpha\gamma^\mu L b_\alpha\cdot
\sum_{q'}e_{q'}\bar q_\beta'\gamma_\mu R q_\beta'\ ,   & O_8 & = &
\frac{3}{2}\bar d_\alpha\gamma^\mu L b_\beta\cdot
\sum_{q'}e_{q'}\bar q_\beta'\gamma_\mu R q_\alpha'\ , \\
O_9 & = & \frac{3}{2}\bar d_\alpha\gamma^\mu L b_\alpha\cdot
\sum_{q'}e_{q'}\bar q_\beta'\gamma_\mu L q_\beta'\ ,   & O_{10} &
= & \frac{3}{2}\bar d_\alpha\gamma^\mu L b_\beta\cdot
\sum_{q'}e_{q'}\bar q_\beta'\gamma_\mu L q_\alpha'\ ,
\label{eq:operators}
\end{array}
\eeq where $\alpha$ and $\beta$ are the $SU(3)$ color indices; $L$
and $R$ are the left- and right-handed projection operators with
$L=(1 - \gamma_5)$, $R= (1 + \gamma_5)$. The sum over $q'$ runs
over the quark fields that are active at the scale $\mu=O(m_b)$,
i.e., $(q'\epsilon\{u,d,s,c,b\})$. The PQCD approach works well
for the leading twist approximation and leading double logarithm
summation. For the Wilson coefficients $C_i(\mu)$
($i=1,\ldots,10$), we will also use the leading order (LO)
expressions, although the next-to-leading order (NLO) calculations
already exist in the literature \cite{buras96}. This is the
consistent way to cancel the explicit $\mu$ dependence in the
theoretical formulae.

For the renormalization group evolution of the Wilson coefficients
from higher scale to lower scale, we use the formulae as given in
Ref.\cite{luy01} directly. At the high $m_W$ scale, the leading
order Wilson coefficients $C_i(M_W)$ are simple and can be found
easily in Ref.\cite{buras96}. In PQCD approach, the scale t may be
larger or smaller than the $m_b$ scale. For the case of $ m_b< t<
m_W$, we evaluate the Wilson coefficients at $t$ scale using
leading logarithm running equations, as given in Eq.(C1) of
Ref.\cite{luy01}. In numerical calculations, we use
$\alpha_s=4\pi/[\beta_1 \ln(t^2/{\Lambda_{QCD}^{(5)}}^2)]$ which
is the leading order expression with $\Lambda_{QCD}^{(5)}=193$MeV,
derived from $\Lambda_{QCD}^{(4)}=250$MeV. Here
$\beta_1=(33-2n_f)/12$, with the appropriate number of active
quarks $n_f$. $n_f=5$ when scale $t$ is larger than $m_b$.

At a given energy scale $t=m_b=4.8$ GeV, the LO Wilson
coefficients $C_i(m_b)$ as given in Ref.\cite{luy01} are \beq
C_1&=& -0.2703, \quad C_2= 1.1188, \quad C_3= 0.0126, \quad C_4=
-0.0270, \non C_5&=& 0.0085, \quad C_6= -0.0326, \quad C_7=
0.0011, \quad C_8= 0.0004, \non C_9&=& -0.0090, \quad C_{10}=
0.0022.
 \label{eq:cimb}
\eeq
If the scale $t < m_b$, then we evaluate the Wilson
coefficients at $t$ scale using the  input of Eq.~(\ref{eq:cimb})
and the formulae in Appendix D of Ref.\cite{luy01} for the case of
$n_f=4$.

\subsection{Wave Functions}\label{ssec:w-f}

In the resummation procedures, the $B$ meson is treated as a
heavy-light system. In general, the B meson light-cone matrix
element can be decomposed as \cite{grozin,bene} \beq
&&\int_0^1\frac{d^4z}{(2\pi)^4}e^{i\bf{k_1}\cdot z}
   \langle 0|\bar{b}_\alpha(0)d_\beta(z)|B(p_B)\rangle \nonumber\\
&=&-\frac{i}{\sqrt{2N_c}}\left\{(\psl_B+m_B)\gamma_5
\left[\phi_B ({\bf k_1})-\frac{\nsl-\vsl}{\sqrt{2}}
\bar{\phi}_B({\bf k_1})\right]\right\}_{\beta\alpha}, \label{aa1}
\eeq
 where $n=(1,0,{\bf 0_T})$, and $v=(0,1,{\bf 0_T})$ are the
unit vectors pointing to the plus and minus directions,
respectively. From the above equation, one can see that there are
two Lorentz structures in the B meson distribution amplitudes.
They obey to the following normalization conditions
 \beq
 \int\frac{d^4 k_1}{(2\pi)^4}\phi_B({\bf k_1})
 =\frac{f_B}{2\sqrt{2N_c}}, ~~~\int \frac{d^4
k_1}{(2\pi)^4}\bar{\phi}_B({\bf k_1})=0.
 \eeq

In general, one should consider these two Lorentz structures in
calculations of $B$ meson decays. However, it can be argued that
the contribution of $\bar{\phi}_B$ is numerically small ~
\cite{luyang,kurimoto}, thus its contribution can be numerically
neglected. In this approximation, we keep minimum number of input
parameters for wave functions. Therefore, we only consider the
contribution of Lorentz structure
\beq
\Phi_B=
\frac{1}{\sqrt{2N_c}} (\psl_B +m_B) \gamma_5 \phi_B ({\bf k_1}),
\label{bmeson}
\eeq
 in our calculation. We use the same wave
functions as in Refs.\cite{luy01,kurimoto,kls01,cl00}.
 In the next section, we will see that
the hard part is always independent of one of the $k_1^+$ and/or
$k_1^-$, if we make some approximations. The B meson wave function
is then the function of  variable $k_1^-$ (or $k_1^+$) and
$k_1^\perp$. \beq \phi_B (k_1^-, k_1^\perp)=\int d k_1^+ \phi
(k_1^+, k_1^-, k_1^\perp). \label{int} \eeq The wave function for
$d\bar{d}$ components in $\pi$ meson are given as \beq
\Phi_{\pi}(P,x,\zeta)\equiv \frac{1}{\sqrt{2N_C}} \left [ \psl
\phi_{\pi}(x)+m_0^{\pi} \phi_{\pi}^{P}(x)+\zeta m_0^{\pi} (\vsl
\nsl - v\cdot n)\phi_{\pi}^{T}(x)\right ] . \eeq

The wave function for $d\bar{d}$ components of $\eta^{(\prime)}$
meson are given as \beq \Phi_{\eta_{d\bar{d}}}(P,x,\zeta)\equiv
\frac{1}{\sqrt{2N_C}} \left [ \psl
\phi_{\eta_{d\bar{d}}}^{A}(x)+m_0^{\eta_{d\bar{d}}}
\phi_{\eta_{d\bar{d}}}^{P}(x)+\zeta m_0^{\eta_{d\bar{d}}} ( \vsl
\nsl - v\cdot n)\phi_{\eta_{d\bar{d}}}^{T}(x) \right ], \eeq where
$P$ and $x$ are the momentum and the momentum fraction of
$\eta_{d\bar{d}}$, respectively. We assumed here that the wave
function of $\eta_{d\bar{d}}$ is same as the $\pi$ wave function.
The parameter $\zeta$ is either $+1$ or $-1$ depending on the
assignment of the momentum fraction $x$. The $s\bar s$ component
of the wave function can be similarly defined.

The transverse momentum $k^\perp$ is usually conveniently
converted to the $b$ parameter by Fourier transformation.
 The initial conditions of leading twist $\phi_i(x)$,
$i=B,\pi,\eta, \eta'$, are of non-perturbative origin, satisfying
the normalization
\beq
\int_0^1\phi_i(x,b=0)dx=\frac{1}{2\sqrt{6}}{f_i}\;, \label{no}
\eeq
with $f_i$ the meson decay constants.

\section{Perturbative Calculations}\label{sec:p-c}

In the previous section we have discussed the wave functions and
Wilson coefficients of the amplitude in eq.(\ref{eq:con1}). In
this section, we will calculate the hard part $H(t)$. This part
involves the four quark operators and the necessary hard gluon
connecting the four quark operator and the spectator quark.  We
will show the whole amplitude for each diagram including wave
functions. Similar to the $B \to \pi \rho$ \cite{ly} and $B \to
\rho \etap$ decays \cite{liu05}, there are 8 type diagrams
contributing to the $B \to \pi \eta^{(\prime)}$ decays, as
illustrated in Figure 1. We first calculate the usual factorizable
diagrams (a) and (b). Operators $O_1$, $O_2$, $O_3$, $O_4$, $O_9$,
and $O_{10}$ are $(V-A)(V-A)$ currents, the sum of their
amplitudes is given as \beq F_{e\pi}&=& 8\pi C_F m_B^4\int_0^1 d
x_{1} dx_{3}\, \int_{0}^{\infty} b_1 db_1 b_3 db_3\,
\phi_B(x_1,b_1) \non & & \times \left\{ \left[(1+x_3) \pp(x_3,
b_3) +(1-2x_3) \rpi (\phi_\pi^p(x_3,b_3)
+\phi_\pi^t(x_3,b_3))\right] \right. \non && \left.\quad  \cdot
\alpha_s(t_e^1)\, h_e(x_1,x_3,b_1,b_3)\exp[-S_{ab}(t_e^1)]
\right.\non && \left. +2\rpi \phi_\pi^p (x_3, b_3)
\alpha_s(t_e^2)h_e(x_3,x_1,b_3,b_1)\exp[-S_{ab}(t_e^2)] \right\},
\label{eq:ab}
 \eeq
where $\rpi=m_0^\pi/m_B$; $C_F=4/3$ is a color
factor.The function $h_e$, the scales $t_e^i$ and the Sudakov
factors $S_{ab}$ are displayed in Appendix \ref{sec:aa}. In the
above equation, we do not include the Wilson coefficients of the
corresponding operators, which are process dependent. They will be
shown later in this section for different decay channels.

\begin{figure}[t,b]
\vspace{-9 cm} \centerline{\epsfxsize=21 cm \epsffile{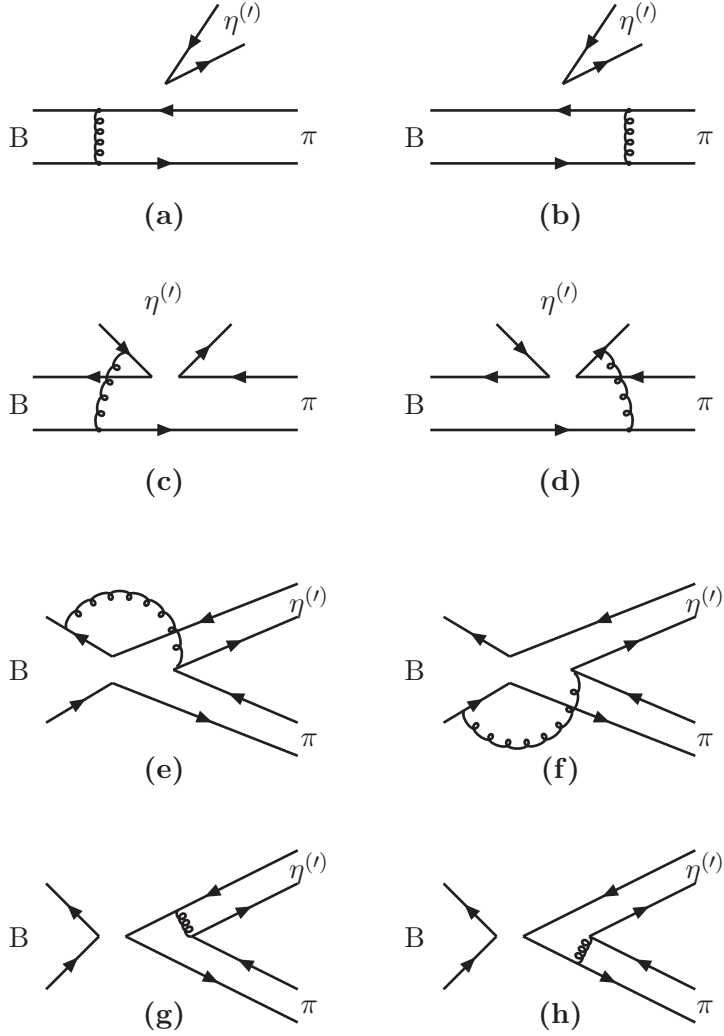}}
\vspace{-7cm} \caption{ Diagrams contributing to the $B\to
\pi\eta^{(\prime)}$
 decays (diagram (a) and (b) contribute to the $B\to \pi$ form
 factor $F_{0,1}^{B\to \pi}$).}
 \label{fig:fig1}
\end{figure}

The form factors of $B$ to $\pi$ decay, $F_{0,1}^{B\to \pi}(0)$,
can thus be extracted from the expression in Eq.~(\ref{eq:ab}),
that is \beq F_0^{B\to\pi}(q^2=0)=F_1^{B\to\pi}(q^2=0)=
F_{e\pi}/m_B^2, \label{eq:f01} \eeq which is identical with that
defined in Ref.\cite{luyang}.

The operators $O_5$, $O_6$, $O_7$, and $O_8$ have a structure of
$(V-A)(V+A)$. In some decay channels, some of these operators
contribute to the decay amplitude in a factorizable way. Since only
the axial-vector part of $(V+A)$ current contribute to the
pseudo-scaler meson production,
$ \langle \pi |V-A|B\rangle \langle \eta |V+A | 0 \rangle = -\langle
\pi |V-A |B  \rangle \langle \eta |V-A|0 \rangle,$ that is
 \beq
 F_{e\pi}^{P1}=-F_{e\pi}\; .
 \eeq
In some other cases, we need to do Fierz
transformation  for these operators to get right color structure
for factorization to work. In this case, we get $(S-P)(S+P)$
operators from $(V-A)(V+A)$ ones. For these $(S-P)(S+P)$ operators,
Fig.~1(a) and 1(b) give
\beq
 F_{e\pi}^{P2}&=& -16\pi C_F m_B^4r_\pi \; \int_{0}^{1}d x_{1}d
x_{3}\,\int_{0}^{\infty} b_1d b_1 b_3d b_3\, \pb(x_1,b_1) \non & &
\times
 \left\{ \left[ \pp(x_3, b_3)+ \rpi((2+x_3) \ppp (x_3, b_3)-x_3\ppt(x_3, b_3)\right]
\right.  \non
& &\left. \cdot \alpha_s (t_e^1)  h_e
(x_1,x_3,b_1,b_3)\exp[-S_{ab}(t_e^1)]\right.  \non & &\left. \
   +\left[x_1\pp(x_3, b_3)-2(x_1-1)\rpi\ppp (x_3, b_3)\right]
\right.  \non
& &\left. \cdot \alpha_s (t_e^2)
 h_e(x_3,x_1,b_3,b_1)\exp[-S_{ab}(t_e^2)] \right\} \; .
 \eeq

For the non-factorizable diagrams 1(c) and 1(d), all three meson
wave functions are involved. The integration of $b_3$ can be
performed using $\delta$ function $\delta(b_3-b_2)$, leaving only
integration of $b_1$ and $b_2$.
For the $(V-A)(V-A)$ operators, the result is
\beq
 M_{e\pi}&=& \frac{16 \sqrt{6}}{3}\pi C_F m_B^4
\int_{0}^{1}d x_{1}d x_{2}\,d x_{3}\,\int_{0}^{\infty} b_1d b_1
b_2d b_2\, \pb(x_1,b_1) \pe(x_2,b_2) \non
 & &\times x_3
 \left[2\rpi \ppt(x_3,b_2)-  \phi_\pi (x_3,b_2)\right
] \non
 & & \cdot \alpha_s(t_f) h_f(x_1,x_2,x_3,b_1,b_2)\exp[-S_{cd}(t_f)]   \; .
\eeq

For the $(V-A)(V+A)$ operators the formulae are different. Here we
have two kinds of contributions from $(V-A)(V+A)$ operators. $M_{e\pi}^{P1}$ and $M_{e\pi}^{P2}$ is for
the $(V-A)(V+A)$ and $(S-P)(S+P)$ type operators respectively:
\beq
M_{e\pi}^{P1}&=&0 , \\
M_{e\pi}^{P2} &=& - M_{e\pi}
  \; . \eeq

The factorizable annihilation diagrams (g) and (h) involve only
$\pi$ and $\etap$ wave functions. There are also three kinds of
decay amplitudes for these two diagrams. $F_{a\pi}$ is for
$(V-A)(V-A)$ type operators, $F_{a\pi}^{P1}$ is for $(V-A)(V+A)$
type operators, while $F_{a\pi}^{P2}$ is for $(S-P)(S+P)$ type
operators:
\beq
F_{a\pi}^{P1}=F_{a\pi}&=& 8\pi C_F m_B^4\int_{0}^{1}dx_{2}\,d
x_{3}\, \int_{0}^{\infty} b_2d b_2b_3d b_3 \, \left\{ \left[x_3
\pp(x_3,b_3) \pe(x_2,b_2)\right.\right.\non &&\left.\left.+2 \re
\rpi((x_3+1)\ppp(x_3,b_3)+(x_3-1) \ppt(x_3,b_3))
\pep(x_2,b_2)\right] \right. \non && \left. \quad \cdot
\alpha_s(t_e^3) h_a(x_2,x_3,b_2,b_3)\exp[-S_{gh}(t_e^3)] \right.
\non && \left. -\left[ x_2 \pp(x_3,b_3) \pe(x_2,b_2)
\right.\right.\non && \left. \left. \quad +2 \re \rpi
\ppp(x_3,b_3)((x_2+1)\pep(x_2,b_2)+(x_2-1) \pet(x_2,b_2) )\right]
\right. \non &&\left. \quad \cdot \alpha_s(t_e^4)
 h_a(x_3,x_2,b_3,b_2)\exp[-S_{gh}(t_e^4)]\right \}\; ,
\eeq
\beq F_{a\pi}^{P2}&=& 16 \pi C_F m_B^4 \int_{0}^{1}d x_{2}\,d
x_{3}\,\int_{0}^{\infty} b_2d b_2b_3d b_3 \,\non &&\times \left\{
\left[x_3 \rpi (\ppp(x_3, b_3)-\ppt(x_3, b_3))\pe(x_2,b_2)+2\re
\pp(x_3,b_3) \pep(x_2,b_2) \right]\right.
 \non
&&\left.\times \alpha_s(t_e^3)
h_a(x_2,x_3,b_2,b_3)\exp[-S_{gh}(t_e^3)]\right.
 \non
 &&\left.+\left[2\rpi
\ppp(x_3,b_3)\pe(x_2,b_2)+x_2\re(\pep(x_2,b_2)-\pet(x_2,b_2))\pp(x_3,b_3)\right]
\right.\non &&\left.\times
 \alpha_s(t_e^4)
 h_a(x_3,x_2,b_3,b_2)\exp[-S_{gh}(t_e^4)]\right\}\; . \eeq

For the non-factorizable annihilation diagrams (e) and (f), again
all three wave functions are involved. Here we have two kinds of
contributions. $M_{a\pi}$, $M_{a\pi}^{P1}$ and $M_{a\pi}^{P2}$
describe the contributions from the $(V-A)(V-A)$, $(V-A)(V+A)$ and
$(S-P)(S+P)$ type operators, respectively,
 \beq
  M_{a\pi}&=&
\frac{16\sqrt{6}}{3}\pi C_F m_B^4\int_{0}^{1}d x_{1}d x_{2}\,d
x_{3}\,\int_{0}^{\infty} b_1d b_1 b_2d b_2\, \phi_B(x_1,b_1)\non
&& \times \left \{ -\left \{x_2 \pp(x_3, b_2) \pe(x_2,b_2)
\right.\right.\non
 & & \left.\left.
 +\rpi\re \left [ (x_2+x_3+2) \pep(x_2,b_2)
 +(x_2-x_3)\pet(x_2,b_2) \right ] \ppp(x_3,b_2)
\right.\right.\non
 & & \left.\left.
  +\left[ (x_2-x_3)\ppp(x_3,b_2) +(x_2+x_3-2)\pet(x_2,b_2)\right ]\ppt(x_3,b_2)
\right\} \right. \non
&& \left.
\quad \cdot \alpha_s(t_f^3)h_f^3(x_1,x_2,x_3,b_1,b_2)\exp[-S_{ef}(t_f^3)]
\right.  \non
&& \left.
  + \left \{ x_3\pp(x_3,b_2) \pe(x_2,b_2) \right.\right. \non
 && \left. \left.
 -\rpi \re \left [ \ppp(x_3,b_2)
 \left [ -(x_2+x_3)\pep(x_2,b_2)+ (x_2-x_3)\pet(x_2,b_2) \right ]
\right. \right. \right.\non
&& \left.\left. \left.
 -\ppt(x_3,b_2)\left [(x_3-x_2)\pep(x_2,b_2)+(x_2+x_3)\pet(x_2,b_2) \right ]
 \right ] \right \}
\right. \non && \left.
 \quad \cdot  \alpha_s(t_f^4)h_f^4(x_1,x_2,x_3,b_1,b_2)\exp[-S_{ef}(t_f^4) ]
 \right \}\; ,
 \eeq
 \beq
M_{a\pi}^{P1}&=& \frac{16\sqrt{6}}{3}\pi C_F m_B^4\int_{0}^{1}d x_{1}d x_{2}\,d x_{3}\,\int_{0}^{\infty} b_1d
b_1 b_2d b_2\, \phi_B(x_1,b_1)
 \non
& &\times \left \{ \left[(x_3-2)
\rpi\pe(x_2,b_2)(\ppp(x_3,b_2)+\ppt(x_3,b_2))-(x_2-2)\re\pp(x_3,b_2)
\right.\right. \non && \left.\left.
(\pep(x_2,b_2)+\pet(x_2,b_2)) \right]
\cdot  \alpha_s(t_f^3)h_f^3(x_1,x_2,x_3,b_1,b_2)\exp[-S_{ef}(t_f^3)]
\right. \non && \left.
 -\left[x_3\rpi\pe(x_2,b_2)(\ppp(x_3, b_2)+\ppt(x_3,
b_2))\right. \right.\non && \left. \left.
 -x_2 \re\pp(x_3,b_2)(\pep(x_2,b_2)+\pet(x_2,b_2))\right]
\right. \non &&
\left.  \cdot
\alpha_s(t_f^4)h_f^4(x_1,x_2,x_3,b_1,b_2)\exp[-S_{ef}(t_f^4)] \right \}
 \; ,
\eeq
\beq
 M_{a\pi}^{P2}&=& \frac{16\sqrt{6}}{3}\pi C_F m_B^4\;
 \int_{0}^{1}d x_{1}d x_{2}\,d x_{3}\,\int_{0}^{\infty} b_1d
b_1 b_2d b_2\, \phi_B(x_1,b_1)\non &&
 \times \left \{
 \left[ x_3 \pp(x_3, b_2) \pe(x_2,b_2)+\rpi\re(\ppp(x_3,b_2)
( (2+x_2+x_3)\pep(x_2,b_2)\right.\right.
 \non
 & & \left. \left.-(x_2-x_3)\pet)+\ppt(-(x_2-x_3)\pep
+(-2+ x_2+x_3)\pet))\right] \right.
\non
&& \left. \cdot\alpha_s(t_f^3)
h_f^3(x_1,x_2,x_3,b_1,b_2)\exp[-S_{ef}(t_f^3)]
 \right. \non
 && \left. +\left[ -x_2 \pp(x_3, b_2)\pe(x_2,b_2)
-\rpi\re((x_2 + x_3)\pep(x_2,b_2)\right. \right. \non
&& \left. \left.
-(x_2 - x_3)\pet(x_2,b_2)\ppp(x_3, b_2) \right.\right. \non
&& \left. \left.
+( (x_3 - x_2)\pep(x_2,b_2) - ( x_2 + x_3)\pet(x_2,b_2))
\ppt(x_3,b_2)) \right] \right. \non
&& \left.
  \cdot \alpha_s(t_f^4)
 h_f^4(x_1,x_2,x_3,b_1,b_2)\exp[-S_{ef}(t_f^4)] \right \}\; . \label{eq:mapip2}
\eeq

In the above equations, we have assumed that $x_1 <<x_2,x_3$.
Since the light quark momentum fraction $x_1$ in $B$ meson is
peaked at the small region, while quark momentum fraction $x_2$ of
$\etap$ is peaked around $0.5$, this is not a bad approximation.
The numerical results also show that this approximation makes very
little difference in the final result. After using this
approximation, all the diagrams are functions of $k_1^-= x_1
m_B/\sqrt{2}$ of B meson only, independent of the variable of
$k_1^+$. Therefore the integration of eq.(\ref{int}) is performed
safely.

If we exchange the $\pi$ and $\etap$ in Fig.~1, the corresponding
expressions of amplitudes for new diagrams will be similar with
those as given in Eqs.(\ref{eq:ab}-\ref{eq:mapip2}), since the
$\pi$ and $\etap$ are all pseudoscalar mesons and have the similar
wave functions. The expressions of amplitudes for new diagrams can
be obtained by the replacements
\beq
\pp \to \pe, \quad \ppp \to
\pep , \quad \ppt \to  \pet, \quad \rpi \to \re.
\eeq
For example, we  find that
\beq
F_{a\eta^{(\prime)}}&=& -F_{a\pi}, \non
F_{a\eta^{(\prime)}}^{P1}&=& -F_{a\pi}^{P1}, \non
F_{a\eta^{(\prime)}}^{P2}&=&F_{a\pi}^{P2}\;.
\eeq
where the form factors $F_{a\eta^{(\prime)}}$, $F_{a\eta^{(\prime)}}^{P1}$ and
$F_{a\eta^{(\prime)}}^{P2}$ describe the contributions induced by the $(V-A)(V-A)$,
$(V-A)(V+A)$ and $(S-P)(S+P)$ operators, respectively.


Before we put the things together to write down the decay amplitudes for the
studied decay modes, we give a brief discussion about the $\eta-\eta^\prime$ mixing
and the gluonic component of the $\eta^\prime$ meson.

The $\eta$ and $\eta^\prime$  are neutral pseudoscalar ($J^P=0^-$) mesons,
and usually
considered as mixtures of the $SU(3)_F$ singlet $\eta_1$ and the octet $\eta_8$:
\beq
\left(\begin{array}{c}
     \eta \\ \eta^{\prime} \end{array} \right)
= \left(\begin{array}{cc}
 \cos{\theta_p} & -\sin{\theta_p} \\
 \sin{\theta_p} & \cos{\theta_p} \\ \end{array} \right)
 \left(\begin{array}{c}
 \eta_8 \\ \eta_1 \end{array} \right),
\label{eq:e-ep} \eeq with \beq \eta_8&=&\frac{1}{\sqrt{6}}\left (
u\bar{u}+d\bar{d}-2s\bar{s}\right ),\non
\eta_1&=&\frac{1}{\sqrt{3}}\left (u\bar{u}+d\bar{d}+s\bar{s}\right
), \label{eq:e1-e8} \eeq where $\theta_p$ is the mixing angle to
be determined by various related experiments \cite{pdg04}. From
previous studies, one obtains the mixing angle $\theta_p$ between
$-20^{\circ}$ to $-10^{\circ}$. One best fit result as given in
Ref.\cite{ekou01} is $-17^{\circ}\leq \theta_p \leq -10^{\circ}$.

As shown in Eqs.~(\ref{eq:e-ep},\ref{eq:e1-e8}),  $\eta$ and
$\eta^\prime$ are generally considered as a linear combination of
light quark pairs. But it should be noted that the $\eta^\prime$
meson may has a gluonic component in order to interpret the
anomalously large branching ratios of $B\to K \eta^\prime$ and
$J/\Psi \to \eta^\prime \gamma$ \cite{ekou01,ekou02}. In
Refs.\cite{rosner83,ekou01,ekou02}, the physical states $\eta$ and
$\eta^\prime$ were defined as
\beq
|\eta> &=& X_\eta \left |
\frac{u\bar{u} + \bar{d}d}{\sqrt{2}} \right > + Y_\eta | s
\bar{s}>, \non |\eta^\prime> &=& X_{\eta^\prime} \left |
\frac{u\bar{u} + \bar{d}d}{\sqrt{2}} \right > + Y_{\eta^\prime} |
s \bar{s}> + Z_{\eta^\prime} |gluonium>, \non \eeq
where
$X_{\etap}, Y_{\etap}$ and $Z_{\eta'}$ parameters describe the
ratios of $u\bar{u}+d \bar{d}$, $s \bar{s}$ and gluonium
($
SU(3)_F$ singlet) componetnt of $\etap$, respectively. In
Ref.\cite{ekou01}, the author shows
that the gluonic admixture in
$\eta^\prime$ can be as large as $26\%$, i.e.
\beq
Z_{\eta'}/\left
( X_{\etap}+Y_{\etap}+Z_{\eta'}\right ) \leq 0.26.
\eeq
According
to paper \cite{ekou02}, a large SU(3) singlet contribution can
help us to explain the large branching ratio for $B \to K
\eta^\prime$ decay, but also result in a large branching ratio for
$B \to K^0 \eta$ decay, $Br(B \to K^0 \eta) \sim 7.0 (13) \times
10^{-6}$ for $\theta_P =-20^\circ (-10^\circ)$ as given in Table
II of Ref.\cite{ekou02}, which is clearly too large than currently
available upper limits \cite{hfag}: \beq Br(B \to K^0 \eta) < 1.9
\times 10^{-6}. \eeq

Although a lot of studies have been done along this direction, but we currently still
do not understand the anomalous  $gg-\eta^\prime$ coupling clearly, and do not know
how to calculate reliably the contributions induced by the gluonic component of
$\eta^\prime$ meson. In this paper, we firstly assume that $\eta^\prime$ does not have
the gluonic component,  and set the quark content of $\eta$ and $\eta^\prime$
as described by Eqs.~(\ref{eq:e-ep},\ref{eq:e1-e8}).
We will also discuss the effects of a non-zero gluonic admixture of $\eta^\prime$ in
next section.


Combining the contributions from different diagrams, the total
decay amplitude for $B^+ \to \pi^+ \eta$ decay can be written as
\beq \sqrt{3} {\cal M}(\pi^+ \eta) &=& F_{e\pi} \left \{\left[
\xi_u \left( C_1 + \frac{1}{3}C_2\right)\right.\right.\non
&&\left.\left. -\xi_t
\left(\frac{7}{3}C_3+\frac{5}{3}C_4-2C_5-\frac{2}{3}C_6
-\frac{1}{2}C_7-\frac{1}{6}C_8 +\frac{1}{3}C_9-\frac{1}{3}
C_{10}\right)\right ] f_\eta^d F_1(\theta_p)\right.
 \non
& &\left. ~-~
\xi_t\left(C_3+\frac{1}{3}C_4-C_5-\frac{1}{3}C_6+\frac{1}{2}C_7
+\frac{1}{6}C_8-\frac{1}{2}C_9-\frac{1}{6}C_{10}\right) f_\eta^s
F_2(\theta_p)\right \} \non && - F_{e\pi}^{P_2} \xi_t \left
(\frac{1}{3}C_5+C_6
-\frac{1}{6}C_7-\frac{1}{2}C_{8}\right)f_{\eta}^d F_1(\theta_p)
\non && + M_{e\pi}\left \{ \left [ \xi_uC_2-\xi_t \cdot
\left(C_3+2C_4-\frac{1}{2}C_9 +\frac{1}{2}C_{10}\right)\right ]
F_1(\theta_p) \right. \non && \left. -\xi_t \left ( C_4
-\frac{1}{2}C_{10}\right )F_2(\theta_p) \right \} \non && -
M_{e\pi}^{P_2}\, \xi_t\,
\left[\left(2C_6+\frac{1}{2}C_8\right)F_1(\theta_p)
+\left(C_6-\frac{1}{2}C_8\right)F_2(\theta_p)\right] \non &&
+\left (M_{a\pi}+M_e+M_a \right )\left [\xi_u C_1 - \xi_t (
C_3+C_9 )\right] F_1(\theta_p)\non && -\left (
M_{a\pi}^{P_1}+M_a^{P_1} \right ) \, \xi_t \,(C_5+  C_7 ) \cdot
F_1(\theta_p) \non && +\left (
F_{a\pi}^{P_2}+F_e^{P_2}+F_a^{P_2}\right ) (-\xi_t)
\left(C_6+C_8+\frac{1}{3}C_5+\frac{1}{3}C_7\right)
F_1(\theta_p)\non &&
 + F_e \cdot\left \{ \left [\xi_u \left(\frac{1}{3}C_1+C_2\right)-
\xi_t \left(\frac{1}{3}C_3+ C_4 +\frac{1}{3}C_9 + C_{10}
\right)\right ] F_1(\theta_p) \right \} , \label{eq:m1} \eeq where
$\xi_u = V_{ub}^*V_{ud}$, $\xi_t = V_{tb}^*V_{td}$, and \beq
F_1(\theta_p)&=& -\sin \theta_p + \cos \theta_p/\sqrt{2},\non
F_2(\theta_p)&=& -\sin \theta_p -\sqrt{2}\cos \theta_p,
\label{eq:f1f2} \eeq are the mixing factors. The Wilson
coefficients $C_i$ should be calculated at the appropriate scale
$t$ using equations as given in the Appendices of
Ref.~\cite{luy01}.

Similarly, the decay amplitude for $B^0 \to \pi^0 \eta$ can be
written as \beq \sqrt{6}{\cal M}(\pi^0 \eta) &=& F_e \left [ \xi_u
\left(C_1+ \frac{1}{3}C_2\right) - \xi_t \left (-\frac{1}{3}C_3
-C_4-\frac{3}{2}C_7-\frac{1}{2}C_8+\frac{5}{3}C_9
+C_{10}\right)\right ] F_1(\theta_p) \non && - F_{e\pi}\left [
 \xi_u \left ( C_1+\frac{1}{3}C_2 \right ) \cdot f_\eta^d F_1(\theta_p)\right. \non
&& \left.
-\xi_t\left(\frac{7}{3}C_3+\frac{5}{3}C_4-2C_5-\frac{2}{3}C_6
-\frac{1}{2}C_7-\frac{1}{6}C_8 +\frac{1}{3}C_9-\frac{1}{3}
C_{10}\right)\cdot f_\eta^d F_1(\theta_p)\right. \non && \left. -
\xi_t\left(C_3+\frac{1}{3}C_4-C_5-\frac{1}{3}C_6+\frac{1}{2}C_7
+\frac{1}{6}C_8-\frac{1}{2}C_9-\frac{1}{6}C_{10}\right) \cdot
f_\eta^s F_2(\theta_p) \right ] \non && - M_{e\pi}\left \{ \left
[\xi_u C_2 -\xi_t \left(C_3 + 2 C_4-\frac{1}{2}C_9
+\frac{1}{2}C_{10}\right)\right ] \cdot F_1(\theta_p)\right. \non
&& \left. -\xi_t \left( C_4-\frac{1}{2}C_{10}\right )
F_2(\theta_p)\right \} \non && -
M_{e\pi}^{P_2}\left[-\xi_t\left(2C_6+\frac{1}{2}C_8\right)F_1(\theta_p)
-\xi_t\left(C_6 -\frac{1}{2}C_8\right)  F_2(\theta_p)\right]\non
&& -(F_{a\pi}^{P_2}+F_{a}^{P_2}+F_e^{P_2})\left[-\xi_t\left(
C_6+\frac{1}{3}C_5-\frac{1}{2}C_8-\frac{1}{6}C_7\right)\right]
\cdot F_1(\theta_p)\non && + \left (M_{a\pi}+M_a+M_e\right) \left
[\xi_u C_2-\xi_t\left( -C_3+\frac{1}{2}C_9
+\frac{3}{2}C_{10}\right)\right ] F_1(\theta_p)\non
 &&
 + \left( M_a^{P_1}+M_{a\pi}^{P_1}\right)
 \; \xi_t \;\left (C_5-\frac{1}{2}C_7  \right ) F_1(\theta_p)
\non &&
-\frac{3}{2}\left(M_{a\pi}^{P_2}+M_{a}^{P_2}+M_e^{P_2}\right)\xi_tC_8
F_1(\theta_p)
  \non && + F_{e\pi}^{P_2}  \; \xi_t \left(\frac{1}{3}C_5+
C_6-\frac{1}{6}C_7 -\frac{1}{2}C_8 \right) \cdot F_1(\theta_p) .
\label{eq:m2} \eeq

The decay amplitudes for $B \to \pi^+ \eta^{\prime}$ and $B \to
\pi^0 \eta'$ can be obtained easily from Eqs.(\ref{eq:m1}) and
(\ref{eq:m2}) by the following replacements
\beq
f_\eta^{d},\; f_\eta^s &\longrightarrow& f_{\eta^\prime}^d, \; f_{\eta^\prime}^s,
\non
F_1(\theta_p) &\longrightarrow & F'_1(\theta_p) =
\cos{\theta_p} + \frac{\sin{\theta_p}}{\sqrt{2}}, \non
F_2(\theta_p) &\longrightarrow & F'_2(\theta_p) = \cos{\theta_p} -
\sqrt{2} \sin{\theta_p}.
\eeq
Note that the possible gluonic component of $\eta'$ meson has been neglected here.

\section{Numerical results and Discussions}\label{sec:n-d}

\subsection{Input parameters and wave functions}

We use the following input parameters in the numerical
calculations \beq
 \Lambda_{\overline{\mathrm{MS}}}^{(f=4)} &=& 250 {\rm MeV}, \quad
 f_\pi = 130 {\rm MeV}, \quad f_B = 190 {\rm MeV}, \non
 m_0^{\eta_{d\bar{d}}}&=& 1.4 {\rm GeV},\quad
 m_0^{\eta_{s\bar{s}}} = 2.4 {\rm GeV}, \quad f_K = 160  {\rm MeV}, \non
 M_B &=& 5.2792 {\rm GeV}, \quad M_W = 80.41{\rm GeV}.
 \label{para}
\eeq
   For the CKM matrix elements, here we adopt the wolfenstein
parametrization for the CKM matrix up to $\mathcal{O}(\lambda^3)$,
\begin{equation}
V_{CKM}= \left(           \begin{array}{ccc}
          1-\frac{\lambda^2}{2} & \lambda & A \lambda^3 (\rho-i \eta)\\
          -\lambda & 1-\frac{\lambda^2}{2} & A \lambda^2 \\
          A \lambda^3 (1-\rho-i \eta )&-A \lambda^2 & 1
          \end{array} \right) , \label{eq:vckm}
\end{equation}
with the parameters $\lambda=0.22, A=0.853, \rho=0.20$ and $\eta=0.33$.

For the $B$ meson wave function, we adopt the model
\beq \phi_B(x,b) &=& N_B x^2(1-x)^2 \mathrm{exp} \left
 [ -\frac{M_B^2\ x^2}{2 \omega_{b}^2} -\frac{1}{2} (\omega_{b} b)^2\right],
 \label{phib}
\eeq where $\omega_{b}$ is a free parameter and we take
$\omega_{b}=0.4\pm 0.05$ GeV in numerical calculations, and
$N_B=91.745$ is the normalization factor for $\omega_{b}=0.4$.
This is the same wave functions as in
Refs.\cite{luy01,kurimoto,kls01,cl00}, which is a best fit for
most of the measured hadronic B decays.

For the light meson wave function, we neglect the $b$ dependant
part, which is not important in numerical analysis. We choose the
wave function of $\pi$ meson  \cite{ball3}: \begin{eqnarray}
 \phi_\pi^A(x) &=&  \frac{3}{\sqrt{6} }
  f_\pi  x (1-x)  \left[1+0.44C_2^{3/2} (2x-1) +0.25 C_4^{3/2}
  (2x-1)\right],\label{piw1}\\
 \phi_{\pi}^P(x) &=&   \frac{f_\pi}{2\sqrt{6} }
   \left[ 1+0.43 C_2^{1/2} (2x-1) +0.09 C_4^{1/2} (2x-1) \right]  ,\\
 \phi_{\pi}^t(x) &=&  \frac{f_\pi}{2\sqrt{6} } (1-2x)
   \left[ 1+0.55  (10x^2-10x+1)  \right]  .    \label{piw}
 \end{eqnarray}
 The Gegenbauer polynomials are defined by
 \begin{equation}
 \begin{array}{ll}
 C_2^{1/2} (t) = \frac{1}{2} (3t^2-1), & C_4^{1/2} (t) = \frac{1}{8}
 (35t^4-30t^2+3),\\
 C_2^{3/2} (t) = \frac{3}{2} (5t^2-1), & C_4^{3/2} (t) = \frac{15}{8}
 (21t^4-14t^2+1).
 \end{array}
 \end{equation}

For $\eta$ meson's wave function, $\phi_{\eta_{d\bar{d}}}^A$,
$\phi_{\eta_{d\bar{d}}}^P$ and $\phi_{\eta_{d\bar{d}}}^T$
represent the axial vector, pseudoscalar and tensor components of
the wave function respectively, for which we utilize the result
 from the light-cone sum rule \cite{ball} including twist-3
contribution:
\beq
\phi_{\eta_{d\bar{d}}}^A(x)&=&\frac{3}{\sqrt{2N_c}}f_xx(1-x)
\left\{ 1+a_2^{\eta_{d\bar{d}}}\frac{3}{2}\left [5(1-2x)^2-1
\right ]\right. \non &&\left. + a_4^{\eta_{d\bar{d}}}\frac{15}{8}
\left [21(1-2x)^4-14(1-2x)^2+1 \right ]\right \},  \non
\phi^P_{\eta_{d\bar{d}}}(x)&=&\frac{1}{2\sqrt{2N_c}}f_x \left \{
1+ \frac{1}{2}\left (30\eta_3-\frac{5}{2}\rho^2_{\eta_{d\bar{d}}}
\right ) \left [ 3(1-2x)^2-1 \right] \right.  \non
&& \left. +
\frac{1}{8}\left
(-3\eta_3\omega_3-\frac{27}{20}\rho^2_{\eta_{d\bar{d}}}-
\frac{81}{10}\rho^2_{\eta_{d\bar{d}}}a_2^{\eta_{d\bar{d}}} \right
) \left [ 35 (1-2x)^4-30(1-2x)^2+3 \right ] \right\} ,  \non
\phi^T_{\eta_{d\bar{d}}}(x) &=&\frac{3}{\sqrt{2N_c}}f_x(1-2x) \non
 && \cdot \left [ \frac{1}{6}+(5\eta_3-\frac{1}{2}\eta_3\omega_3-
\frac{7}{20}\rho_{\eta_{d\bar{d}}}^2
-\frac{3}{5}\rho^2_{\eta_{d\bar{d}}}a_2^{\eta_{d\bar{d}}})(10x^2-10x+1)\right
],  \non \eeq with \beq a^{\eta_{d\bar{d}}}_2&=& 0.44, \quad
a^{\eta_{d\bar{d}}}_4=0.25,\quad
 a_1^K=0.20, \quad a_2^K=0.25, \non
\rho_{\eta_{d\bar{d}}}&=&m_{\pi}/{m_0^{\eta_{d\bar{d}}}}, \quad
\eta_3=0.015, \quad \omega_3=-3.0. \eeq

 We assume that the wave function of $u\bar{u}$ is same as the wave
  function of $d\bar{d}$.
For the wave function of the $s\bar{s}$ components, we also use
the same form as $d\bar{d}$ but with $m^{s\bar{s}}_0$ and $f_y$
instead of $m^{d\bar{d}}_0$ and $f_x$, respectively. For $f_x$ and
$f_y$, we use the values as given in Ref.\cite{kf} where isospin
symmetry is assumed for $f_x$ and $SU(3)$ breaking effect is
included for $f_y$:
 \beq
 f_x=f_{\pi}, \ \ \ f_y=\sqrt{2f_K^2-f_{\pi}^2}.\ \ \
\label{eq:7-5} \eeq

These values are translated to the values in the two mixing angle
method, which is often used in vacuum saturation approach as:
\beq
f_8 &=&169  {\rm MeV}, \quad f_1=151  {\rm MeV},  \non
\theta_8&=& -25.9^{\circ} (-18.9^{\circ}), \quad \theta_1=-7.1^{\circ}
(-0.1^{\circ}),
\eeq
where the pseudoscalar mixing angle $\theta_p$ is taken as
$-17^{\circ}$ ($-10^{\circ}$) \cite{ekou01}.
The parameters $m_0^i$
$(i=\eta_{d\bar{d}(u\bar{u})}, \eta_{s\bar{s}})$ are defined as:
\beq
m_0^{\eta_{d\bar{d}(u\bar{u})}}\equiv m_0^\pi \equiv
\frac{m_{\pi}^2}{(m_u+m_d)}, \qquad m_0^{\eta_{s\bar{s}}}\equiv
\frac{2M_K^2-m_{\pi}^2}{(2m_s)}.
 \label{eq:19}
\eeq

We include full expression of twist$-3$ wave functions for light
mesons. The twist$-3$ wave functions are also adopted from QCD sum
rule  calculations \cite{bf}. We will see later that this set of
 parameters will give good results for $B \to \pi
 \eta^{(\prime)}$ decays.
Using the above chosen wave functions and the central values of
relevant input parameters, we find the numerical values of the
corresponding form factors at zero momentum transfer:
 \beq
F_0^{B\to \pi}(q^2=0)&=& 0.30, \non
F_0^{B \to \eta}(q^2=0)&=& 0.15, \non
F_0^{B \to \eta^{\prime}}(q^2=0)&=& 0.14. \label{eq:aff0}
\eeq
These values agree well with those as given in Refs.~\cite{ball,kf,sum}.

\subsection{Branching ratios}

For $B \to \pi \etap$ decays, the decay amplitudes in
Eqs.~(\ref{eq:m1}) and (\ref{eq:m2}) can be rewritten as
 \beq
{\cal M} &=& V_{ub}^*V_{ud} T -V_{tb}^* V_{td} P= V_{ub}^*V_{ud} T
\left [ 1 + z e^{ i ( \alpha + \delta ) } \right], \label{eq:ma}
\eeq
where
\beq
z=\left|\frac{V_{tb}^* V_{td}}{ V_{ub}^*V_{ud} }
\right| \left|\frac{P}{T}\right| \label{eq:zz}
\eeq
is the ratio of penguin to tree contributions, $\alpha = \arg
\left[-\frac{V_{td}V_{tb}^*}{V_{ud}V_{ub}^*}\right]$ is the weak
phase (one of the three CKM angles), and $\delta$ is the relative
strong phase between tree (T) and penguin (P) diagrams. The ratio
$z$ and the strong phase $\delta$ can be calculated in the pQCD
approach. One can leave the CKM angle $\alpha$ as a free parameter
and explore the CP asymmetry parameter dependence on it.

For $B \to \pi^+ \eta$ decay, for example, one can find ``T" and
``P" terms by comparing the decay amplitude as defined in
Eq.~(\ref{eq:m1}) with that in Eq.~(\ref{eq:ma}), \beq
T(\pi^+\eta)&=& \frac{F_1(\theta_p)}{\sqrt{3}} \cdot \left \{
F_{e\pi} \left( C_1 + \frac{1}{3}C_2\right) f_\eta^d +M_{e\pi} C_2
\right. \non && \left.+  F_e \left ( \frac{1}{3}C_1 + C_2\right )
+ \left ( M_a + M_e +M_{a\pi}\right ) C_1 \right \}, \label{eq:tt}
\eeq \beq \sqrt{3} P(\pi^+ \eta) &=& F_{e\pi} \left [
\left(\frac{7}{3}C_3+\frac{5}{3}C_4-2C_5-\frac{2}{3}C_6
-\frac{1}{2}C_7-\frac{1}{6}C_8 +\frac{1}{3}C_9-\frac{1}{3}
C_{10}\right) f_\eta^d F_1(\theta_p)\right.
 \non
& &\left.+
\left(C_3+\frac{1}{3}C_4-C_5-\frac{1}{3}C_6+\frac{1}{2}C_7
+\frac{1}{6}C_8-\frac{1}{2}C_9-\frac{1}{6}C_{10}\right) f_\eta^s
F_2(\theta_p)\right ]
 \non
&& + F_{e\pi}^{P_2} \left (\frac{1}{3}C_5+C_6
-\frac{1}{6}C_7-\frac{1}{2}C_{8}\right)f_{\eta}^d F_1(\theta_p)
\non && - M_{e\pi}\left [ - \left(C_3+2C_4-\frac{1}{2}C_9
+\frac{1}{2}C_{10}\right) F_1(\theta_p)  - \left ( C_4
-\frac{1}{2}C_{10}\right )F_2(\theta_p)\right ]\non &&
+M_{e\pi}^{P_2}\left[\left(2C_6+\frac{1}{2}C_8\right)F_1(\theta_p)
+\left(C_6-\frac{1}{2}C_8\right)F_2(\theta_p)\right]\non &&
+(M_{a\pi}+M_e+M_a)( C_3+C_9 )F_1(\theta_p) +\left (
M_{a\pi}^{P_1}+M_a^{P_1} \right )(C_5+
 C_7 ) \cdot  F_1(\theta_p)\non &&+(F_{a\pi}^{P_2}+F_e^{P_2}+F_a^{P_2})
  F_1(\theta_p)\left(C_6+C_8+\frac{1}{3}C_5+\frac{1}{3}C_7\right)
 \non
 && + F_e\left(\frac{1}{3}C_3+ C_4 +\frac{1}{3}C_9 + C_{10} \right)
F_1(\theta_p) ,\label{eq:pp}\eeq
Similarly, one can obtain the expressions of the corresponding
tree and penguin terms for the remaining three decays.

Using the ``T" and ``P" terms as given in Eqs.(\ref{eq:tt}) and (\ref{eq:pp}),
it is easy to calculate the ratio $z$ and
the strong phase $\delta$ for the decay in study. For  $B^+ \to
\rho^+ \eta$ and $\rho^+ \eta'$ decays, we find numerically that
\beq
z(\pi^+\eta) &=&0.33, \qquad \delta (\pi^+\eta)=-136^\circ , \label{eq:zd1}\\
z(\pi^+\eta^\prime) &=&0.25 , \qquad
\delta(\pi^+\eta^\prime)=-130^\circ .\label{eq:zd2} \eeq The main
errors of the ratio $z$ and the strong phase $\delta$ are induced
by the uncertainty of $\omega_b=0.4 \pm 0.05$ GeV and $m_0^\pi=1.4
\pm 0.1$ GeV and small in magnitude. The reason is
that the errors induced by the uncertainties of input parameters
are largely cancelled in the ratio. We therefore use the central
values of $z$ and $\delta$ in the following numerical
calculations, unless explicitly stated otherwise.

From Eq.~(\ref{eq:ma}), it is easy to write the decay amplitude for the
corresponding charge conjugated decay mode
\beq
\overline{\cal M}
&=& V_{ub}V_{ud}^* T -V_{tb} V_{td}^* P = V_{ub}V_{ud}^* T \left[1
+z e^{i(-\alpha + \delta)} \right]. \label{eq:mb}
 \eeq
Therefore the CP-averaged branching ratio for $B^0 \to \pi \etap$
is
\beq
Br= (|{\cal M}|^2 +|\overline{\cal M}|^2)/2 =  \left|
V_{ub}V_{ud}^* T \right| ^2 \left[1 +2 z\cos \alpha \cos \delta
+z^2 \right], \label{br}
\eeq
where the ratio $z$ and the strong
phase $\delta$ have been defined in Eqs.(\ref{eq:ma}) and
(\ref{eq:zz}). It is easy to see that the CP-averaged branching
ratio is a function of $\cos \alpha$. This gives a potential
method to determine the CKM angle $\alpha$ by measuring only the
CP-averaged branching ratios with PQCD calculations.

Using  the wave functions and the input parameters as specified in
previous sections,  it is straightforward  to calculate the
branching ratios for the four considered decays. The theoretical
predictions in the PQCD approach for the branching ratios of the
decays under consideration are the following \beq Br(\ B^+ \to
\pi^+ \eta) &=& \left [4.1 ^{+1.3}_{-0.9}(\omega_b) ^{+0.4}_{-0.3}
(m_0^\pi) ^{+0.6}_{-0.5} (\alpha )\right ] \times 10^{-6},
 \label{eq:brp-eta}\\
Br(\ B^+ \to \pi^+ \eta^{\prime}) &=& \left [2.4^{+0.8}_{-0.5}(
\omega_b) \pm 0.2 (m_0^\pi) \pm 0.3  (\alpha ) \right ] \times
10^{-6},
\label{eq:brp-etap}\\
Br(\ B^0 \to \pi^0 \eta) &=& \left [0.23
^{+0.04}_{-0.03}(\omega_b) ^{+0.04}_{-0.03} (m_0^\pi)\pm 0.05
(\alpha) \right ]\times 10^{-6},
\label{eq:br0-eta} \\
Br(\ B^0 \to \pi^0 \eta^{\prime}) &=& \left [0.19 \pm
0.02(\omega_b) \pm 0.03(m_0^\pi) ^{+0.04}_{-0.05} (\alpha ) \right
]\times 10^{-6} \label{eq:br0-etap}, \eeq for
$\theta_p=-17^\circ$, and \beq
 Br(\ B^+ \to \pi^+ \eta) &=&\left [3.3 ^{+1.0}_{-0.8}(\omega_b)
\pm 0.3 (m_0^\pi) \pm 0.4(\alpha ) \right ]\times 10^{-6},
\label{eq:brp-eta1}\\
Br(\ B^+ \to \pi^+ \eta^{\prime}) &=& \left [3.2 ^{+1.1}_{-0.7}(
\omega_b) \pm 0.3 (m_0^\pi) \pm 0.4 (\alpha )\right
]\times10^{-6},
\label{eq:brp-etap1} \\
Br(\ B^0 \to \pi^0 \eta) &=&\left [0.17 \pm 0.02(\omega_b) \pm
0.02(m_0^\pi)
^{+0.03}_{-0.04} ( \alpha )\right ])\times 10^{-6}, \label{eq:br0-eta1}\\
 Br(\ B^0 \to \pi^0 \eta^{\prime}) &=&\left [ 0.28
^{+0.04}_{-0.02}(\omega_b) \pm 0.04(m_0^\pi)
 \pm 0.05 (\alpha )\right ] \times 10^{-6}, \label{eq:br0-etap1}
\eeq for $\theta_p=-10^\circ$. The main errors are induced by the
uncertainties of $\omega_b=0.4 \pm 0.04$ GeV, $m_0^\pi = 1.4 \pm
0.1$ GeV and $\alpha =100^\circ \pm 20^\circ$, respectively.

The PQCD predictions for the branching ratios of considered decays
agree  well with the measured values or the upper limits as shown in
Eqs.(\ref{eq:exp1}-\ref{eq:ulimits}). Furthermore, the pQCD predictions also agree
well with the theoretical predictions in the QCDF approach, for example, as given in Ref.~\cite{bn03b}:
\beq
Br(\ B^+ \to \pi^+ \eta) &=&  \left ( 4.7 ^{+2.7}_{-2.3}\right ) \times 10^{-6}, \non
Br(\ B^+ \to \pi^+ \eta^\prime) &=&  \left ( 3.1 ^{+1.9}_{-1.7}\right ) \times 10^{-6}, \non
Br(\ B^0 \to \pi^0 \eta) &=& \left ( 0.28 ^{+0.48}_{-0.28}\right ) \times 10^{-6}, \non
Br(\ B^0 \to \pi^0 \eta^{\prime}) &=&  \left ( 0.17 ^{+0.33}_{-0.17}\right ) \times 10^{-6},
\label{eq:br24}
\eeq
where the individual errors as given in Ref.~\cite{bn03b} have been added
in quadrature.

\begin{figure}[tb]
\centerline{\mbox{\epsfxsize=8cm\epsffile{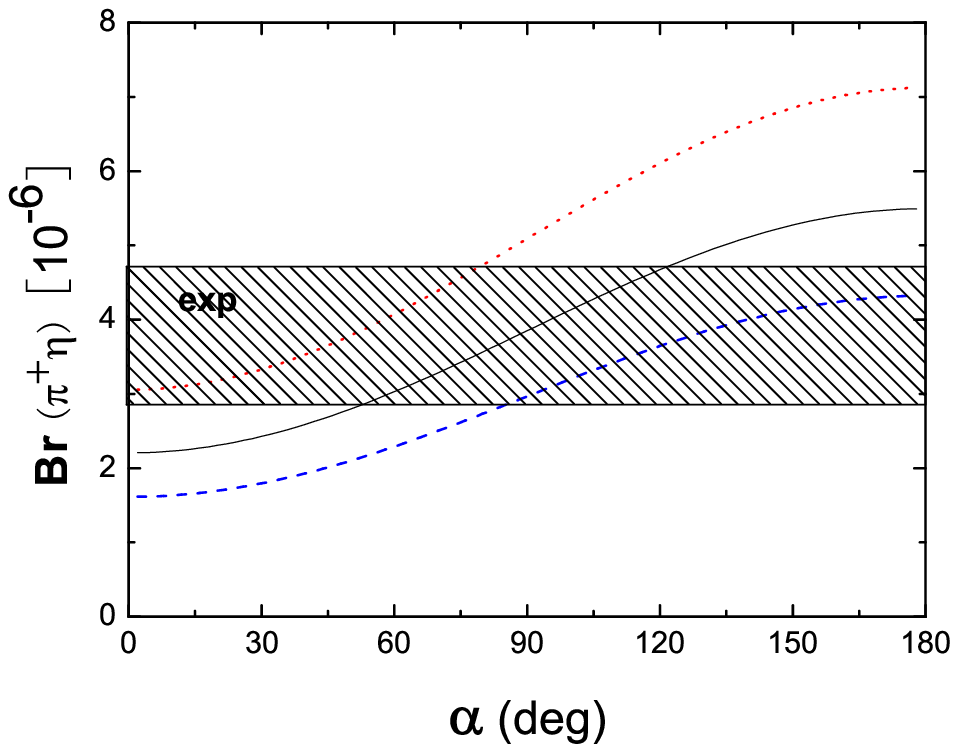}\epsfxsize=8cm\epsffile{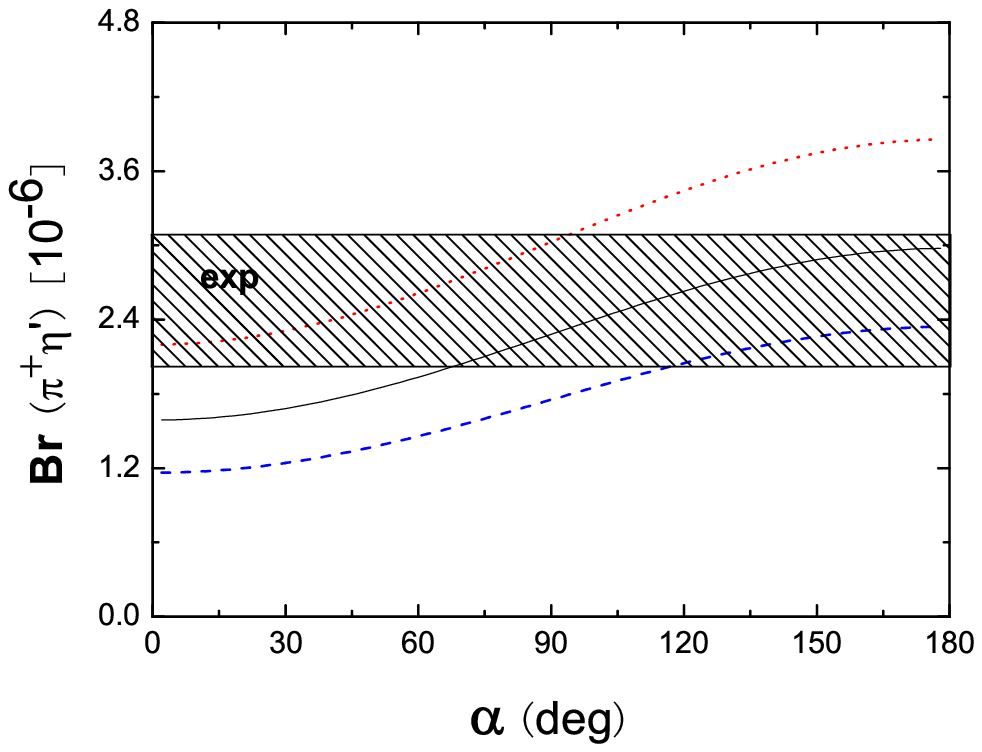}}}
\vspace{0.3cm}
 \caption{The $\alpha$ dependence of the branching ratios (in unit of $10^{-6}$)
 of $B^+\to \pi^+ \eta^{(\prime)}$ decays for $m_0^\pi=1.4$ GeV,
 $\theta_p=-17^\circ$, $\omega_b=0.36 $ GeV
 (dotted curve), $0.40$ GeV (solid  curve) and $0.44$ GeV(dashed curve).
 The gray band shows the experimental data.}
 \label{fig:fig2}
\end{figure}

\begin{figure}[tb]
\centerline{\mbox{\epsfxsize=8cm\epsffile{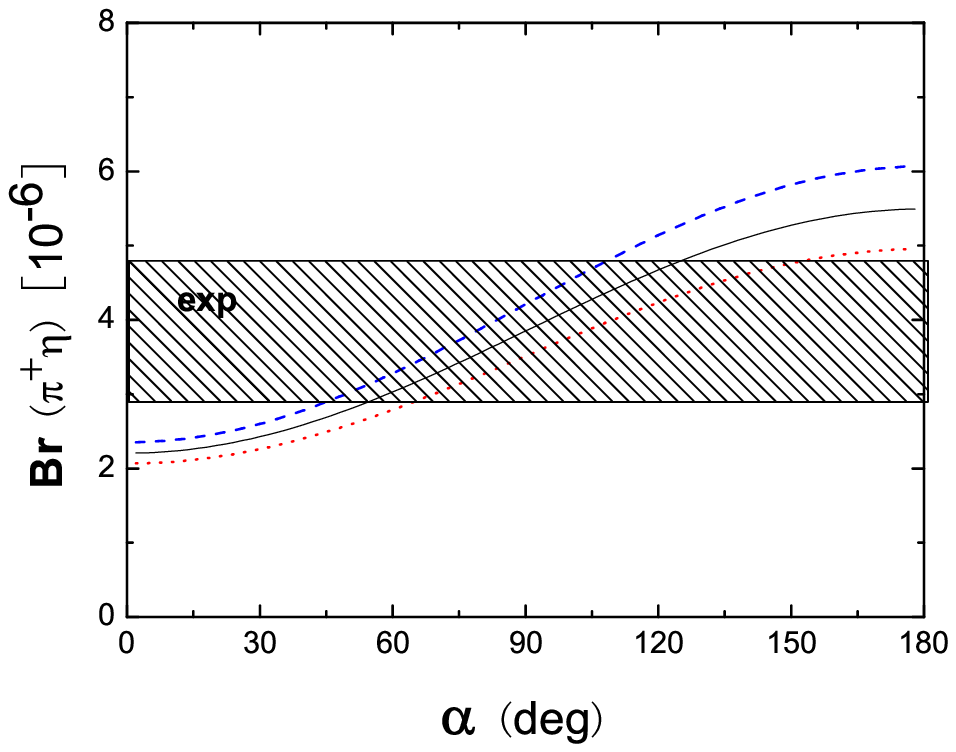}\epsfxsize=8cm\epsffile{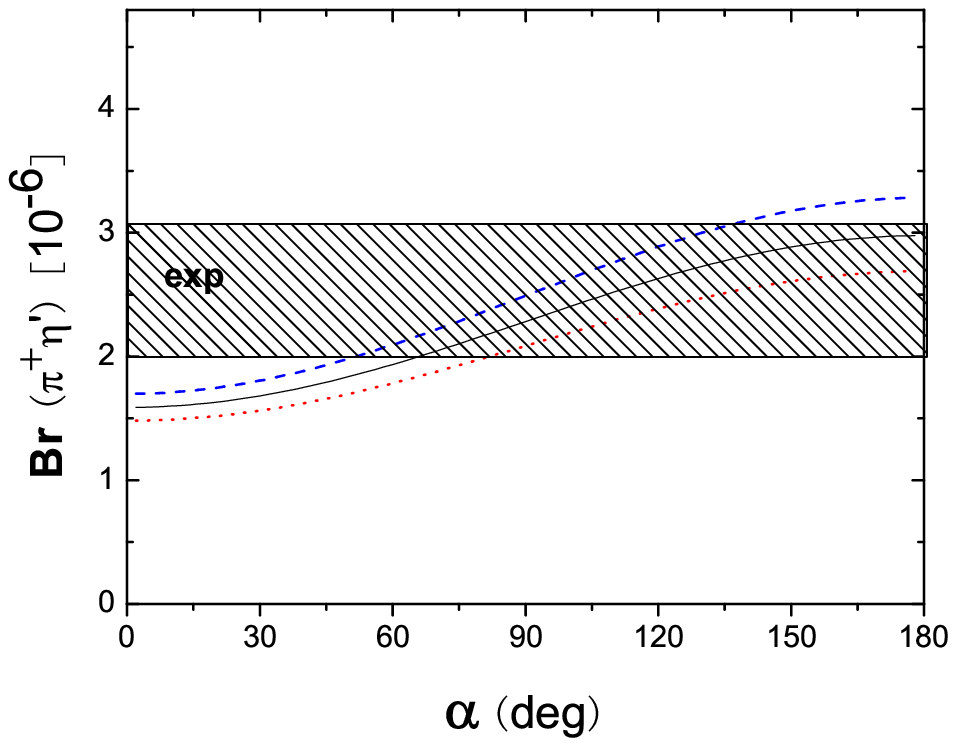}}}
\vspace{0.3cm}
 \caption{The $\alpha$ dependence of the branching ratios (in unit of $10^{-6}$)
 of $B^+\to \pi^+ \eta^{(\prime)}$ decays for $\omega_b=0.4$ GeV,
 $\theta_p=-17^\circ$, $m_0^\pi=1.3$ GeV
 (dotted curve), $1.4$ GeV (solid  curve) and $1.5$ GeV (dashed curve).
 The gray band shows the data.}
 \label{fig:fig3}
\end{figure}

\begin{figure}[tb]
\centerline{\mbox{\epsfxsize=8cm\epsffile{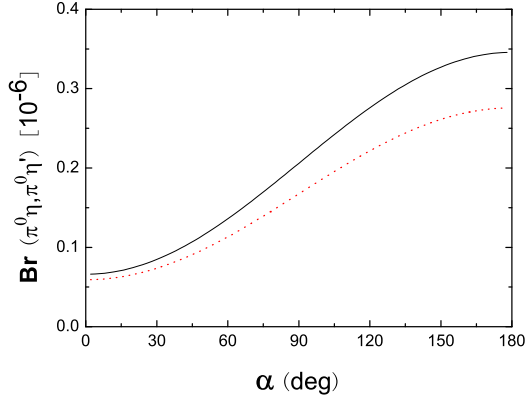}}}
 \caption{The $\alpha$ dependence of the branching ratios (in unit of $10^{-6}$)
 of  $\pi^0 \eta$ (solid  curve) and $\pi^0 \eta^{\prime}$ (dotted curve)  decays
  for $m_0^\pi=1.4$ GeV, $\theta_p=-17^\circ$, $\omega_b=0.40$ GeV .}
 \label{fig:fig4}
\end{figure}

It is worth stressing  that the theoretical predictions in the
PQCD approach have relatively large theoretical errors induced by the still
large uncertainties of many input parameters, such as $\omega_b$,
$m_0^\pi$ and $\theta_p$. In our analysis, we consider the
constraints on these parameters from analysis of other well
measured decay channels. For example, the constraint $1.1
\mbox{GeV} \leq m_0^\pi \leq 1.9 \mbox{GeV}$ was obtained from the
phenomenological studies for $B \to \pi \pi$ decays \cite{luy01},
while the constraint  of $\alpha \approx 100^\circ \pm 20 ^\circ$
were obtained by direct measurements or from the global fit
\cite{hfag,charles05}.
 From numerical calculations, we get to know that the main
errors come from the uncertainty of $\omega_b$, $m_0^\pi$,
$\alpha$ and $\theta_p$.

In Figs.~\ref{fig:fig2} and \ref{fig:fig3}, we present,
respectively, the pQCD predictions of the branching ratios of $B
\to \pi^+ \eta$ and $\pi^+ \eta^\prime$ decays for
$\theta_p=10^\circ$, $\omega_b=0.4\pm 0.04$ GeV, $m_0^\pi=1.4\pm
0.1$ GeV and $\alpha=[0^\circ,180^\circ]$. Fig.~\ref{fig:fig4}
shows  the $\alpha$-dependence of the pQCD predictions of the
branching ratios of $B \to \pi^0 \etap$ decays for
$\theta_p=10^\circ$, $\omega_b=0.4$ GeV, $m_0^\pi=1.4$ GeV and
$\alpha=[0^\circ,180^\circ]$.

From the numerical results and the figures we observe that the
pQCD predictions are sensitive to the variations of
$\omega_b$ and $m_0^\pi$. The parameter $m_0^\pi$ originates from
the chiral perturbation theory and have a value near 1 GeV. The
$m_0^{\pi}$ parameter characterizes the relative size of twist 3
contribution to twist 2 contribution. Because of the chiral
enhancement of $m_0^{\pi}$, the twist 3 contribution become
comparable in size with the twist 2 contribution. The branching
ratios of $Br(B \to \pi \etap)$ are also sensitive to the
parameter $m_0^\pi$, but not as strong as the $\omega_b$
dependence.

\subsection{CP-violating asymmetries }

Now we turn to the evaluations of the CP-violating asymmetries of
$B \to \pi \etap$ decays in pQCD approach. For $B^+ \to \pi^+
\eta$ and $B^+ \to \pi^+ \eta^\prime$ decays, the direct
CP-violating asymmetries $\acp$ can be defined as:
 \beq
\acp^{dir} =  \frac{|\overline{\cal M}|^2 - |{\cal M}|^2}{
 |\overline{\cal M}|^2+|{\cal M}|^2}=
\frac{2 z \sin \alpha \sin\delta}{1+2 z\cos \alpha \cos \delta
+z^2}, \label{eq:acp1}
 \eeq
where the ratio $z$ and the strong phase $\delta$ have been
defined in previous subsection and are calculable in pQCD
approach.

\begin{figure}[tb]
\vspace{-1cm} \centerline{\epsfxsize=10cm \epsffile{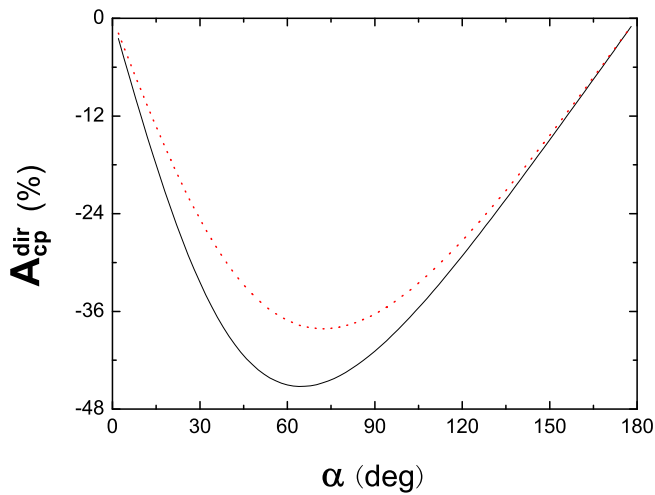}}
\vspace{-0.5cm}
 \caption{The direct CP asymmetries (in percentage) of $B^+\to \pi^+ \eta$ (solid curve)
 and $B^+\to \pi^+ \eta^{\prime}$ (dotted curve) as a function of CKM
angle $\alpha$.}
 \label{fig:fig5}
\end{figure}

It is easy  to calculate the CP-violating asymmetries with $z$ and
$\delta$. In Fig.~\ref{fig:fig5}, we show the $\alpha-$dependence
of the direct CP-violating asymmetries $\acp^{dir}$ for
$B^\pm \to \pi^\pm \eta$ (the solid curve) and $B^\pm \to \pi^\pm
\eta'$ (the dotted curve) decay, respectively. From
Fig.~\ref{fig:fig5}, one can see that the CP-violating asymmetries
$\acp^{dir}(B^\pm \to \pi^\pm \etap)$ are large in magnitude,
about $35\%$ for $\alpha \sim 100^\circ$. So large CP-violating
asymmetry plus large ($\sim 10^{-6}$) branching ratios are
measurable in current B factory experiments.

The pQCD predictions for $\acp^{dir}$ and the major theoretical
errors for $B^\pm \to \pi^\pm \etap$ decays are
\beq
\acp^{dir}(B^\pm \to \pi^\pm \eta) &=& \left ( -37 ^{+8}_{-6}(\alpha) \pm
4(\omega_b)^{+0}_{-1}(m_0^\pi) \right ) \times 10^{-2} \label{eq:acp-a}, \\
\acp^{dir}(B^\pm \to \pi^\pm \eta^\prime) &=& \left ( -33 ^{+6}_{-4}(\alpha)
{^{+4}_{-6}}(\omega_b) {^{+0}_{-2}}(m_0^\pi)  \right ) \times
10^{-2} \label{eq:acp-b},
\eeq
where the dominant errors come from the variations of $\omega_b=0.4\pm 0.04$ GeV,
$m_0^\pi = 1.4 \pm 0.1$ GeV and $\alpha=100^\circ \pm 20^\circ$.

By comparing the above numerical results with those measured values as given in
Eqs.(\ref{eq:acpexp1}) and (\ref{eq:acpexp2})
\footnote{We used the same convention as the BaBar and Belle Collaboration
\cite{babar,belle} to define the CP-violating asymmetry for $B \to \pi \etap$
decays. }, we find that
\begin{enumerate}
\item
The pQCD predictions for the direct CP-violating asymmetry for both
$B \to \pi \eta$ and $\pi \eta^\prime$ decays are large in magnitude and
have a moderate theoretical error because of the cancelation in the ratios.

\item
For $B^\pm  \to \pi^\pm \eta$ decay, the pQCD prediction for $\acp^{dir}$
has the same sign with the measured value and also consistent with it within
$2\sigma$ errors.
For $B^\pm  \to \pi^\pm \eta^\prime$ decay, however, the pQCD prediction for $\acp^{dir}$
has the opposite sign with the measured value.

\item
For $B^\pm \to \pi^\pm \eta^\prime$ decay, although there exist a clear difference
between the pQCD prediction of $\acp^{dir}$ and the
data, but it is too early to draw any reliable information from such difference
because of the still large theoretical and experimental errors.
More theoretical studies (for example, calculation of next-to-leading order
contributions \cite{nlo}) and more accurate measurements are needed
to clarify this discrepancy.

\end{enumerate}

In Ref.~\cite{bn03b}, by using the ``default values" of input parameters,
the authors presented their predictions for
$\acp^{dir}(B \to \pi^\pm \etap)$ in the QCDF approach
\beq
\acp^{dir}(B^\pm \to \pi^\pm \eta) &=& \left ( -14.9^{+4.9\; +8.3\; +1.3\; +17.4}_{
-5.4\; -7.4\; -0.8\; -17.3} \right ) \times 10^{-2} \label{eq:acp-1}, \\
\acp^{dir}(B^\pm \to \pi^\pm \eta^\prime) &=& \left ( -8.6^{+2.8\;
+10.5\; +0.7\; +20.4}_{ -3.1\; -9.0\;\;\;  -0.7\; -20.4} \right )
\times 10^{-2} \label{eq:acp-2}, \eeq where the first error comes
from the variation of the CKM parameters, the second error refers
to the variation of  $\mu\sim m_b$, quark masses, decay constants,
form factors, and the mixing angle $\theta_p$. The third error
corresponds to the uncertainty due to the Gegenbauer moments in
the expansion of the LCDAs. The last but largest error is induced
by the uncertainty of the unknown annihilation contributions. In
fact, these numbers are similar to those generalized factorization
approach \cite{akl2}, since the mechanism of strong phase is the
same for these two approaches.

By using the ``set $S_4$"  input parameters \cite{bn03b}, the central values of the
QCDF  prediction can become positive simultaneously
\beq
\acp^{dir}(B^\pm \to \pi^\pm \eta) &=&  5.6 \times 10^{-2} \label{eq:acp-1b}, \\
\acp^{dir}(B^\pm \to \pi^\pm \eta^\prime) &=& 11.1 \times 10^{-2} \label{eq:acp-2b}.
\eeq

 From the numerical results as given in Eqs.(\ref{eq:acp-1}-\ref{eq:acp-2b}), one can
see that the dominant theoretical error from the annihilation
contributions in QCD factorization approach are too large to make
any meaningful comparisons between the theoretical predictions and
the data for $\acp^{dir}(B \to \pi \etap)$. The reason is that the
annihilation contributions play a key role for producing the
strong phase of the two-body charmless B meson decays, but
unfortunately they are incalculable in QCD factorization approach.

We now study the CP-violating asymmetries for $B^0 \to \pi^0
\etap$ decays. For these neutral decay modes, the effects of
$B^0-\bar{B}^0$ mixing should be considered. For $B^0$ meson
decays, we know that $\Delta \Gamma/\Delta m_d \ll 1$ and $\Delta
\Gamma/\Gamma \ll 1$. The CP-violating asymmetry of $B^0(\bar B^0)
\to \pi^0 \eta^{(\prime)}$ decay is time dependent and can be
defined as
\beq A_{CP} &\equiv& \frac{\Gamma\left (\overline{B_d^0}(\Delta t)
\to f_{CP}\right)
- \Gamma\left(B_d^0(\Delta t) \to f_{CP}\right )}{ \Gamma\left
(\overline{B_d^0}(\Delta t) \to f_{CP}\right )
+ \Gamma\left (B_d^0(\Delta t) \to f_{CP}\right ) }\non
&=& A_{CP}^{dir} \cos
(\Delta m  \Delta t) + A_{CP}^{mix} \sin (\Delta m  \Delta t),
\label{eq:acp-def}
\eeq
where $\Delta m$ is the mass difference between the two $B_d^0$ mass
eigenstates, $\Delta t =t_{CP}-t_{tag} $ is the time difference between the
tagged $B^0$ ($\overline{B}^0$) and the accompanying $\overline{B}^0$ ($B^0$)
with opposite b flavor decaying to the final CP-eigenstate
$f_{CP}$ at the time $t_{CP}$. The direct and mixing induced
CP-violating asymmetries $A_{CP}^{dir}$ and $A_{CP}^{mix}$ can be
written as
\beq
\acp^{dir}=\frac{ \left | \lambda_{CP}\right |^2 -1 } {1+|\lambda_{CP}|^2},
\qquad
A_{CP}^{mix}=\frac{ 2Im (\lambda_{CP})}{1+|\lambda_{CP}|^2},
\label{eq:acp-dm}
\eeq
where the CP-violating parameter $\lambda_{CP}$ is
\beq
\lambda_{CP} = \frac{ V_{tb}^*V_{td} \langle \pi^0 \etap |H_{eff}|
\overline{B}^0\rangle} { V_{tb}V_{td}^* \langle \pi^0 \etap |H_{eff}|
B^0\rangle} = e^{2i\alpha}\frac{ 1+z e^{i(\delta-\alpha)} }{
1+ze^{i(\delta+\alpha)} }.
\label{eq:lambda2}
\eeq
Here the ratio $z$ and the strong phase $\delta$ have been defined previously. In
PQCD approach, since both $z$ and $\delta$ are calculable, it is
easy to find the numerical values of $A_{CP}^{dir}$ and
$A_{CP}^{mix}$ for the considered decay processes.

\begin{figure}[tb]
\vspace{-1cm} \centerline{\epsfxsize=10cm \epsffile{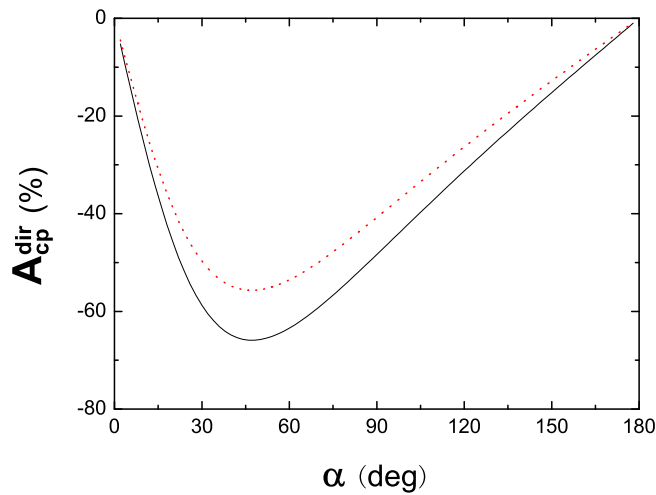}}
\vspace{-0.5cm}
 \caption{The direct CP asymmetry $A_{CP}^{dir}$ (in percentage) of
 $B^0\to \pi^0 \eta$ (solid curve)
 and $B^0\to \pi^0 \eta^{\prime}$ (dotted curve) as a function of CKM
angle $\alpha$.} \label{fig:fig6}
\end{figure}

\begin{figure}[htb]
\vspace{-1cm} \centerline{\epsfxsize=10 cm \epsffile{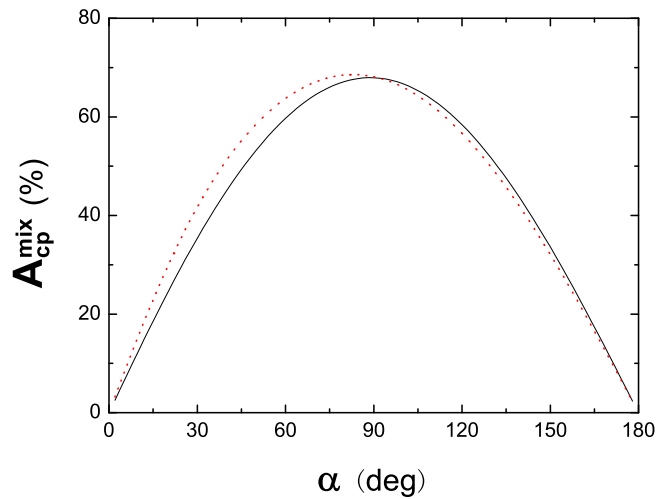}}
\vspace{-0.5cm} \caption{The mixing induced CP asymmetry
$A_{CP}^{mix}$ (in percentage) of $B^0\to \pi^0 \eta$ (solid
curve) and $B^0\to \pi^0 \eta^{\prime}$ (dotted curve) as a
function of CKM angle $\alpha$ .} \label{fig:fig7}
\end{figure}

In Figs.~\ref{fig:fig6} and \ref{fig:fig7}, we show the
$\alpha-$dependence of the direct and the mixing-induced
CP-violating asymmetry $A_{CP}^{dir}$ and $A_{CP}^{mix}$ for $B^0
\to \pi^0 \eta$ (solid curve) and $B^0 \to \pi^0 \eta^\prime$
(dotted curve) decays, respectively.

The pQCD predictions for the direct and mixing induced
CP-violating asymmetries of $B^0 \to \pi^0 \etap$ decays are
\beq
\acp^{dir}(B^0 \to \pi^0 \eta) &=& \left ( -42 ^{+9}_{-12}(\alpha)
^{+3}_{-2}(\omega_b) ^{+1}_{-3}(m_0^\pi)
  \right ) \times 10^{-2} \label{eq:acp-d1}, \\
\acp^{dir}(B^0 \to \pi^0 \eta^\prime) &=& \left ( -36 ^{+10}_{-9}(\alpha)
^{+2}_{-1}(\omega_b) ^{+2}_{-3}(m_0^\pi) \right )
\times 10^{-2} \label{eq:acp-d2},  \\
\acp^{mix}(B^0 \to \pi^0 \eta) &=& \left ( 67 ^{+0}_{-9}(\alpha)
^{+5}_{-6}(\omega_b) ^{+1}_{-2}(m_0^\pi)
\right ) \times 10^{-2} \label{eq:acp-m1}, \\
\acp^{mix}(B^0 \to \pi^0 \eta^\prime) &=& \left ( 67 ^{+0}_{-9}(\alpha)
^{+4}_{-6}(\omega_b) ^{+1}_{-3}(m_0^\pi)  \right ) \times
10^{-2} \label{eq:acp-m2},
\eeq
where the dominant errors come from the variations of $\omega_b=0.4\pm 0.05$ GeV,
$m_0^\pi=1.4\pm 0.1$ GeV  and $\alpha=100^\circ \pm 20^\circ$.

As a comparison, we present the QCDF predictions for $\acp^{dir}(B^0 \to \pi^0 \etap)$
directly quoted from Ref.~\cite{bn03b}
\beq
\acp^{dir}(B^0 \to \pi^0 \eta) &=&
\left ( -17.9^{+5.2\; +7.9\; +1.2\; +33.4}_{
-4.1\; -14.1\; -1.4\; -32.9} \right ) \times 10^{-2}, \\
\acp^{dir}(B^0 \to \pi^0 \eta^\prime) &=&
\left ( -19.2^{+5.5\; +7.7\; +4.1\; +35.7}_{
-4.3\; -7.8\; -3.3 \; -35.8} \right ) \times 10^{-2}\label{eq:acp-db},
\eeq
where the ``default values" of the input parameters have been used \cite{bn03b}, and
the error sources are the same as those for the numerical results in
Eqs.~(\ref{eq:acp-1}) and (\ref{eq:acp-2}). Currently, no relevant experimental
measurements for the CP-violating asymmetries of $B \to \pi^0 \etap$ decays
are available. For the direct CP-violating asymmetries of $B^0 \to \pi^0 \etap$ decays,
the theoretical predictions in pQCD and QCDF approach have the same sign, but
the theoretical errors are clearly too large to make a meaningful comparison.
One has to wait for the improvements in both the experimental measurements and the calculation
of high order contributions.

\begin{figure}[htb]
\vspace{-1cm} \centerline{\epsfxsize=10 cm \epsffile{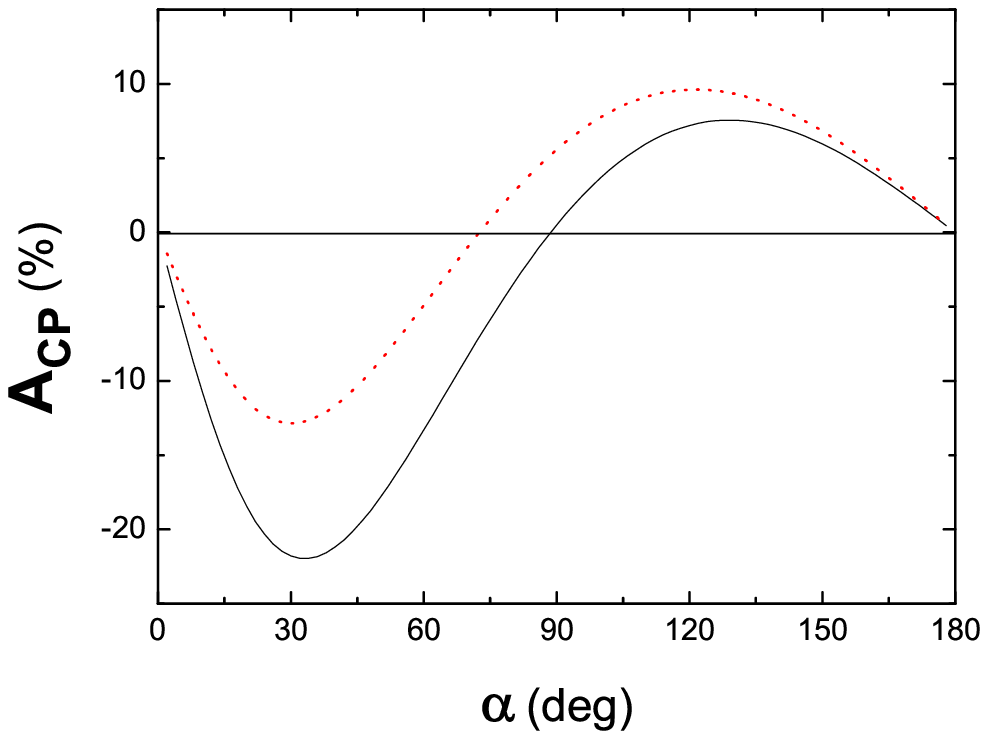}}
\vspace{-0.5cm} \caption{The total CP asymmetry $A_{cp}^{tot}$ (in
percentage) of $B^0\to \pi^0 \eta$ (solid curve) and $B^0\to \pi^0
\eta^{\prime}$ (dotted curve) as a function of CKM angle $\alpha$
.} \label{fig:fig8}
\end{figure}

If we integrate the time variable $t$, we will get the total CP
asymmetry for $B^0 \to \pi^0 \etap$ decays,
\beq
A_{CP}=\frac{1}{1+x^2} A_{CP}^{dir} + \frac{x}{1+x^2} A_{CP}^{mix},
\eeq
where $x=\Delta m/\Gamma=0.771$ for the $B^0-\overline{B}^0$
mixing \cite{pdg04}. In Fig.\ref{fig:fig8},
we show the $\alpha$-dependence of the total CP asymmetry $A_{CP}$
for $B^0 \to \pi^0 \eta$ (solid curve) and $B^0 \to \pi^0
\eta^\prime$ (solid curve) decay, respectively. Numerically, we
found
\beq
\acp^{tot}(B^0 \to \pi^0 \eta)&=& \left(
+3.7^{+3.5}_{-7.3} (\alpha) ^{+4.6}_{-3.9}(\omega_b) ^{+0.3}_{-1.0}(m_0^\pi)
\right ) \times 10^{-2}, \\
\acp^{tot}(B^0 \to \pi^0 \eta^\prime)&=& \left(+7.7 ^{+1.9}_{-5.1}(\alpha)
^{+0.3}_{-0.8}(\omega_b)^{+1.9}_{-5.1}(m_0^\pi)  \right ) \times 10^{-2}.
\eeq

\subsection{Effects of possible gluonic component of $\eta^\prime$}

Up to now, we have not considered the possible  contributions to
the branching ratios and CP-violating asymmetries of $B \to \pi
\eta^\prime$ decays induced by the possible gluonic component of
$\eta^\prime$ \cite{rosner83,ekou01,ekou02}. When $Z_{\eta^\prime}
\neq 0$, a decay amplitude ${\cal M}'$ will be produced by the
gluonic component of $\eta^\prime$. Such decay amplitude may
construct or destruct with the ones from the $q\bar{q}$
($q=u,d,s$) components of $\eta^\prime$, the branching ratios of
the decays in question may be increased or decreased accordingly.

Unfortunately, we currently do not know how to calculate this kind
of contributions reliably. But we can treat it as an theoretical
uncertainty. For $|M'/M(q\bar{q})| \sim 0.1-0.2$, for example, the
resulted uncertainty for the branching ratios as given in
Eqs.(\ref{eq:brp-etap}) and (\ref{eq:br0-etap}) will be around
twenty to thirty percent.

 From Eq.~(\ref{eq:exp2},\ref{eq:brp-etap}), one can see that the theoretical result of
$Br(B^+ \to \pi^+ \eta^\prime)$ in the PQCD approach agree very
well with the measured values within one standard deviation.
Furthermore, the pQCD predictions for the branching ratios of $B
\to \rho \etap$ decays also show very good agreement with the data
\cite{liu05}. We therefore believe that the gluonic admixture of
$\eta^\prime$ should be small, and most possibly not as important
as expected before.

As for the CP-violating asymmetries of $B \to \pi \eta^\prime$
decays, the possible contributions of the gluonic components of
the $\eta^\prime$ meson are largely cancelled in the ratio.

\section{summary }

In this paper,  we calculate the branching ratios and CP-violating
asymmetries of $B^0 \to \pi^0 \eta$, $B^0 \to \pi^0
\eta^{\prime}$, $B^+ \to \pi^+ \eta$, and $B^+ \to \pi^+
\eta^{\prime}$ decays in the PQCD factorization approach.

Besides the usual factorizable diagrams, the non-factorizable and
annihilation diagrams are also calculated analytically. Although
the non-factorizable and annihilation contributions are
sub-leading for the branching ratios of the considered decays, but
they are not negligible. Furthermore these diagrams provide the
necessary strong phase required by a non-zero CP-violating
asymmetry for the considered decays.

From our calculations and phenomenological analysis, we found the following
results:
\begin{itemize}

\item
From analytical calculations, the form factors for $B \to
\eta$, $B \to \eta^\prime$ and $B\to \pi$ transitions can be
extracted. The PQCD results for these form factors are $F_{0,1}^{B\to
\pi}(0)=0.30$, $F_{0,1}^{B\to \eta}(0)=0.15$ and $F_{0,1}^{B\to
\eta^\prime}(0)=0.14$.

\item
For the CP-averaged branching ratios of the four considered decay modes,
the theoretical predictions in PQCD approach are
\beq
Br(B^+ \to \pi^+ \eta ) &=& \left ( 4.1 ^{+1.5}_{-1.1}\right ) \times 10^{-6}, \non
Br(B^+ \to \pi^0 \eta^{(\prime)}) &=&\left ( 2.4 ^{+0.9}_{-0.6}\right )
\times 10^{-6}, \non
Br(B^0 \to \pi^0 \eta ) &=&\left ( 0.23 \pm 0.08\right ) \times 10^{-6}, \non
Br(B^+ \to \pi^0 \eta^{(\prime)}) &=&\left ( 0.19 \pm 0.05 \right )
\times 10^{-6},
\eeq
where the various errors as specified in Eqs.~(\ref{eq:brp-eta}-\ref{eq:br0-etap}) have been added in
quadrature. Although the theoretical uncertainties are still large
(can reach $40\%$), the leading PQCD predictions agree very well
with the measured values or currently available experimental upper
limits, and are also well consistent with the results obtained by
employing the QCD factorization approach.

\item
For the CP-violating asymmetries, the theoretical
predictions in PQCD approach are
\beq
A_{CP}^{dir}(B^\pm \to \pi^\pm \eta) & = &
\left ( -37^{+9}_{-7}\right ) \times 10^{-2}, \non
A_{CP}^{dir}(B^\pm \to \pi^\pm \eta^{\prime}) & =&
\left ( -33^{+7}_{-8}\right ) \times 10^{-2},\non
A_{CP}^{dir}(B^0 \to \pi^0 \eta) & = &
\left ( -42 ^{+10}_{-13}\right ) \times 10^{-2}, \non
A_{CP}^{mix}(B^0 \to \pi^0 \eta)  &=&
\left ( +67 ^{+5}_{-11}\right ) \times 10^{-2}, \non
A_{CP}^{dir}(B^0 \to \pi^0 \eta^{\prime}) & = &
\left ( -36^{+11}_{-10}\right ) \times 10^{-2},\non
A_{CP}^{mix}(B^0 \to \pi^0 \eta^\prime) & =&
\left ( +67 ^{+5}_{-11}\right ) \times 10^{-2},
\eeq
where the various errors as specified in Eqs.~(\ref{eq:acp-a},\ref{eq:acp-b}) and
(\ref{eq:acp-d1}-\ref{eq:acp-m2}) have been added in quadrature. The uncertainties
are around twenty to forty percent.
And finally, the time integrated CP asymmetry for the neutral decay modes are
\beq
\acp^{tot}(B^0 \to \pi^0 \eta)&=& \left(+3.7^{+5.8}_{-8.3} \right ) \times 10^{-2}, \non
\acp^{tot}(B^0 \to \pi^0 \eta^\prime)&=&
\left(+7.7 ^{+2.7}_{-5.4} \right ) \times 10^{-2}.
\eeq

\item
For $B^\pm \to \pi^\pm \eta$ decay, the pQCD prediction for $\acp^{dir}$ has the same sign with the
measured value and consistent with the data within two standard deviations.
For $B^\pm  \to \pi^\pm \eta^\prime$ decay, however, the pQCD prediction for $\acp^{dir}$
has an opposite sign with the measured value. Great improvements in both the
theoretical calculations and experimental measurements are needed to clarify this discrepancy.

\item
From the very good agreement of the pQCD predictions for  the CP-averaged branching
ratios $Br(B^+ \to \pi^+ \etap) $ and $Br(B^+ \to \rho^+ \etap) $ \cite{liu05}
with the measured values, we believe that the gluonic admixture of
$\eta^\prime$ should be small, and most possibly not as important
as expected before.

\end{itemize}

\begin{acknowledgments}

We are very grateful to Li Ying for helpful discussions. This work
is partly supported  by the National Natural Science Foundation of
China under Grant No.10275035, 10475085, 10575052, and by the
Research Foundations of Jiangsu Education Committee and Nanjing
Normal University under Grant No.~214080A916 and 2003102TSJB137.

\end{acknowledgments}


\begin{appendix}

\section{Related Functions }\label{sec:aa}

We show here the function $h_i$'s, coming from the Fourier
transformations  of $H^{(0)}$,
\beq
 h_e(x_1,x_3,b_1,b_3)&=&
 K_{0}\left(\sqrt{x_1 x_3} m_B b_1\right)
 \left[\theta(b_1-b_3)K_0\left(\sqrt{x_3} m_B
b_1\right)I_0\left(\sqrt{x_3} m_B b_3\right)\right.
 \non
& &\;\left. +\theta(b_3-b_1)K_0\left(\sqrt{x_3}  m_B b_3\right)
I_0\left(\sqrt{x_3}  m_B b_1\right)\right] S_t(x_3), \label{he1}
\eeq
 \beq
 h_a(x_2,x_3,b_2,b_3)&=&
 K_{0}\left(i \sqrt{x_2 x_3} m_B b_3\right)
 \left[\theta(b_3-b_2)K_0\left(i \sqrt{x_3} m_B
b_3\right)I_0\left(i \sqrt{x_3} m_B b_2\right)\right.
 \non
& &\;\;\;\;\left. +\theta(b_2-b_3)K_0\left(i \sqrt{x_3}  m_B
b_2\right) I_0\left(i \sqrt{x_3}  m_B b_3\right)\right] S_t(x_3),
\label{he3} \eeq
 \beq
 h_{f}(x_1,x_2,x_3,b_1,b_2) &=&
 \biggl\{\theta(b_2-b_1) \mathrm{I}_0(M_B\sqrt{x_1 x_3} b_1)
 \mathrm{K}_0(M_B\sqrt{x_1 x_3} b_2)
 \non
&+ & (b_1 \leftrightarrow b_2) \biggr\}  \cdot\left(
\begin{matrix}
 \mathrm{K}_0(M_B F_{(1)} b_1), & \text{for}\quad F^2_{(1)}>0 \\
 \frac{\pi i}{2} \mathrm{H}_0^{(1)}(M_B\sqrt{|F^2_{(1)}|}\ b_1), &
 \text{for}\quad F^2_{(1)}<0
\end{matrix}\right),
\label{eq:pp1}
 \eeq
\beq
h_f^3(x_1,x_2,x_3,b_1,b_2) &=& \biggl\{\theta(b_1-b_2)
\mathrm{K}_0(i \sqrt{x_2 x_3} b_1 M_B)
 \mathrm{I}_0(i \sqrt{x_2 x_3} b_2 M_B)+(b_1 \leftrightarrow b_2) \biggr\}
 \non
& & \cdot
 \frac{\pi i}{2} \mathrm{H}_0^{(1)}(\sqrt{x_1+x_2+x_3-x_1 x_3-x_2 x_3}\ b_1 M_B),
 \label{eq:pp4}
\eeq
 \beq
 h_f^4(x_1,x_2,x_3,b_1,b_2) &=&
 \biggl\{\theta(b_1-b_2) \mathrm{K}_0(i \sqrt{x_2 x_3} b_1 M_B)
 \mathrm{I}_0(i \sqrt{x_2 x_3} b_2 M_B)
 \non
&+& (b_1 \leftrightarrow b_2) \biggr\} \cdot \left(
\begin{matrix}
 \mathrm{K}_0(M_B F_{(2)} b_1), & \text{for}\quad F^2_{(2)}>0 \\
 \frac{\pi i}{2} \mathrm{H}_0^{(1)}(M_B\sqrt{|F^2_{(2)}|}\ b_1), &
 \text{for}\quad F^2_{(2)}<0
\end{matrix}\right), \label{eq:pp3}
\eeq
where $J_0$ is the Bessel function and  $K_0$, $I_0$ are
modified Bessel functions $K_0 (-i x) = -(\pi/2) Y_0 (x) + i
(\pi/2) J_0 (x)$, and $F_{(j)}$'s are defined by
\beq
F^2_{(1)}&=&(x_1 -x_2) x_3\;,\\
F^2_{(2)}&=&(x_1-x_2) x_3\;\;.
 \eeq

The threshold resummation form factor $S_t(x_i)$ is adopted from
Ref.\cite{kurimoto}
\beq S_t(x)=\frac{2^{1+2c} \Gamma
(3/2+c)}{\sqrt{\pi} \Gamma(1+c)}[x(1-x)]^c,
\eeq
where the parameter $c=0.3$. This function is normalized to unity.

The Sudakov factors used in the text are defined as
\beq
S_{ab}(t) &=& s\left(x_1 m_B/\sqrt{2}, b_1\right) +s\left(x_3 m_B/\sqrt{2},
b_3\right) +s\left((1-x_3) m_B/\sqrt{2}, b_3\right) \non
&&-\frac{1}{\beta_1}\left[\ln\frac{\ln(t/\Lambda)}{-\ln(b_1\Lambda)}
+\ln\frac{\ln(t/\Lambda)}{-\ln(b_3\Lambda)}\right],
\label{wp}\\
S_{cd}(t) &=& s\left(x_1 m_B/\sqrt{2}, b_1\right)
 +s\left(x_2 m_B/\sqrt{2}, b_2\right)
+s\left((1-x_2) m_B/\sqrt{2}, b_2\right) \non
 && +s\left(x_3
m_B/\sqrt{2}, b_1\right) +s\left((1-x_3) m_B/\sqrt{2}, b_1\right)
\non
 & &-\frac{1}{\beta_1}\left[2
\ln\frac{\ln(t/\Lambda)}{-\ln(b_1\Lambda)}
+\ln\frac{\ln(t/\Lambda)}{-\ln(b_2\Lambda)}\right],
\label{Sc}\\
S_{ef}(t) &=& s\left(x_1 m_B/\sqrt{2}, b_1\right)
 +s\left(x_2 m_B/\sqrt{2}, b_2\right)
+s\left((1-x_2) m_B/\sqrt{2}, b_2\right) \non
 && +s\left(x_3
m_B/\sqrt{2}, b_2\right) +s\left((1-x_3) m_B/\sqrt{2}, b_2\right)
\non
 &
&-\frac{1}{\beta_1}\left[\ln\frac{\ln(t/\Lambda)}{-\ln(b_1\Lambda)}
+2\ln\frac{\ln(t/\Lambda)}{-\ln(b_2\Lambda)}\right],
\label{Se}\\
S_{gh}(t) &=& s\left(x_2 m_B/\sqrt{2}, b_1\right)
 +s\left(x_3 m_B/\sqrt{2}, b_2\right)
+s\left((1-x_2) m_B/\sqrt{2}, b_1\right) \non
 &+& s\left((1-x_3)
m_B/\sqrt{2}, b_2\right)
-\frac{1}{\beta_1}\left[\ln\frac{\ln(t/\Lambda)}{-\ln(b_1\Lambda)}
+\ln\frac{\ln(t/\Lambda)}{-\ln(b_2\Lambda)}\right], \label{ww}
\eeq
where the function $s(q,b)$ are defined in the Appendix A of
Ref.\cite{luy01}. The scale $t_i$'s in the above equations are
chosen as
\beq
t_{e}^1 &=& {\rm max}(\sqrt{x_3} m_B,1/b_1,1/b_3)\;,\\
t_{e}^2 &=& {\rm max}(\sqrt{x_1}m_B,1/b_1,1/b_3)\;,\\
t_{e}^3 &=& {\rm max}(\sqrt{x_3}m_B,1/b_2,1/b_3)\;,\\
t_{e}^4 &=& {\rm max}(\sqrt{x_2}m_B,1/b_2,1/b_3)\;,\\
t_{f} &=& {\rm max}(\sqrt{x_1 x_3}m_B, \sqrt{(x_1-x_2) x_3}m_B,1/b_1,1/b_2)\;,\\
t_{f}^3 &=& {\rm max}(\sqrt{x_1+x_2+x_3-x_1 x_3-x_2 x_3}m_B,
    \sqrt{x_2 x_3} m_B,1/b_1,1/b_2)\;,\\
t_{f}^4 &=&{\rm max}(\sqrt{x_2 x_3} m_B,1/b_1,1/b_2)\; .
\eeq

\end{appendix}



\newpage

\begin{thebibliography}{99}

\bibitem{cpv}
I.I.~Bigi and A.I.~Sanda, {\it CP Violation}  (Cambridge
University Press, Cambridge, England, 2000);
G.C.~Branco, L.~Lavoura and J.P.~Silva, {\it CP Violation} (Oxford University
Press, Oxford, England, 1999);
R.~Fleischer, \pr {\bf 370} (2002) 537; T.~Hurth, \rmp {\bf 75} (2003) 1159.

\bibitem{bbns99}
M.~Beneke, G.~Buchalla, M.~Neubert, and C.T.~Sachrajda, \prl 83 (1999) 914.


\bibitem{cl97}
C.-H. V.~Chang and H.-n.~Li,\prd {\bf 55} (1997) 5577;
T.-W.~Yeh and H.-n.~Li, \prd {\bf 56} (1997) 1615.

\bibitem{li2003}
H.-n.~Li, Prog. Part. $\&$ Nucl. Phys. {\bf 51} (2003) 85, and
reference therein.

\bibitem{lb80}
G.P.~Lepage and S. Brodsky, \prd {\bf 22} (1980) 2157;
J.~Botts and G.~Sterman, \npb {\bf 325} (1989) 62.


\bibitem{du03}
D.S.~Du, H.J.~Gong, J.F.~Sun, D.S.~Yang, and G.H.~Zhu,
\prd {\bf 65} (2002) 074001, {\it ibid} {\bf 65} (2002) 094025;
J.F.~Sun, G.H.~Zhu and D.S.~Du, \prd {\bf 68} (2003) 054003.

\bibitem{yy01}
M.Z.~Yang and Y.D.~Yang, \npb {\bf 609} (2001) 469;
M.~Beneke and  M.~Neubert, \npb {\bf 651} (2003) 225.

\bibitem{bn03b}
M.~Beneke and  M.~Neubert, \npb {\bf 675}, 333 (2003).

\bibitem{luy01}
C.-D.~L\"u, K.~Ukai and M.Z.~Yang, \prd {\bf 63} (2001) 074009.

\bibitem{kls01}
Y.-Y.~Keum, H.-n.~Li and A.I.~Sanda, \plb {\bf 504} (2001) 6;
 \prd {\bf 63}  (2001) 054008.

\bibitem{li01}
 H.-n. Li, \prd {\bf 64}  (2001) 014019;
 S.~Mishima,\plb {\bf 521} (2001) 252;
 C.-H.~Chen, Y.-Y.~Keum, and H.-n.~Li,\prd {\bf 64} (2001) 112002;
 A.I.~Sanda and K.~Ukai, Prog. Theor. Phys. {\bf 107} (2002) 421;
 C.D.~L\"u,\epjc {\bf 24} (2002) 121;
 C.-H.~Chen,Y.-Y.~Keum, and H.-n.~Li, \prd {\bf 66}  (2002) 054013;
 Y.-Y.~Keum and A.I.~Sanda, \prd {\bf 67}  (2003) 054009.

\bibitem{kklls04}
Y.-Y.~Keum,  T.~Kurimoto, H.-n.~Li, C.D. L\"u, and A.I.~Sanda,
\prd {\bf 69}  (2004) 094018;
Y.~Li and C.D.~L\"u, \jpg {\bf 29} (2003) 2115;
C.D.~L\"u, \prd {\bf 68}  (2003) 097502;
X.Q~Yu, Y.~Li and C.D.~L\"u, \prd {\bf 71} (2005) 074026;
C.D.~L\"u, Y.L.~Shen and J.~Zhu, \epjc 41 (2005) 311;
J.~Zhu, Y.L.~Shen and C.D.~L\"u, \prd 72 (2005) 054015; and hep-ph/0506316;
Y.~Li and C.D.~L\"u, Commun.  Theor. Phys. 44 (2005) 659, hep-ph/0502038;
and hep-ph/0508032;
C.D.~L\"u, M.~Matsumori, A.I.~Sanda, and M.Z.~Yang, \prd 72 (2005) 094005.

\bibitem{li05}
Y.~Li, C.D.~L\"u, Z.J.~Xiao, and X.Q.~Yu, \prd {\bf 70} (2004) 034009;
Y.~Li, C.D.~L\"u, and Z.J.~Xiao, \jpg {\bf 31} (2005) 273.

\bibitem{liu05}
X.~Liu, H.S.~Wang, Z.J.~Xiao, L.B.~Guo, and C.D.~L\"u, hep-ph/0509362.

\bibitem{buras96}
G.~Buchalla, A.J.~Buras, and M.E.~Lautenbacher, Rev. Mod. Phys. {\bf 68}  (1996) 1125.

\bibitem{ali98}
A.~Ali, G.~Kramer, and C.D.~L\"u, \prd {\bf 58} (1998) 094009.

\bibitem{babar}
BaBar Collaboration, B.~Aubert {\it et al.}, \prd {\bf 70} (2004) 032006;
BaBar Collaboration, B.~Aubert {\it et al.}, \prl {\bf 95} (2005) 131801.

\bibitem{belle}
Belle Collaboration, P.~Chang {\it et al.}, \prd {\bf 71} (2005) 091106(R);
Belle Collaboration, K.~Abe {\it et  al.}, hep-ex/0508030 and hep-ex/0509016.

\bibitem{hfag}
Heavy Flavor Averaging Group, http://www.slac.stanford.edu/xorg/hfag.

\bibitem{li02}
H.-n.~Li, \prd {\bf 66} (2002) 094010.

\bibitem{soft}
H.-n.~Li and B.~Tseng,\prd {\bf 57} (1998) 443.

\bibitem{grozin}
 A.G.~Grozin and M.~Neubert, \prd {\bf 55} (1997) 272;
 M.~Beneke and T.~Feldmann, \npb {\bf 592} (2001) 3.

\bibitem{bene}
M.~Beneke, G.~Buchalla, M.~Neubert, and C.T.~Sachrajda,\npb {\bf 591}  (2000) 313.

\bibitem{luyang}
C.D.~L\"u, M.Z.~Yang, \epjc {\bf 28} (2003) 515.

\bibitem{kurimoto}
T.~Kurimoto, H.-n.~Li, A.I.~Sanda, \prd {\bf 65}  (2002) 014007.

\bibitem{cl00}
C.-H.~Chen and H.-n.~Li,\prd {\bf 63} (2000) 014003.


\bibitem{ly}
C.D.~L\"u, M.Z.~Yang, \epjc {\bf 23}  (2002) 275.

\bibitem{pdg04}
Particle Data Group, S.~Eidelman {\it et al.}, \plb {\bf 592} (2004) 1.

\bibitem{ekou01}
E.~Kou, \prd {\bf 63}  (2001) 054027.

\bibitem{ekou02}
E.~Kou and A.I.~Sanda,\plb {\bf 525} (2002) 240.

\bibitem{rosner83}
J.L.~Rosner, \prd {\bf 27} (1983) 1101.

\bibitem{ball3}
V.M.~Braun and I.E.~Filyanov, \zpc {\bf 48}  (1990) 239;
P.~Ball, \jhep 01 (1999) 010.

\bibitem{ball}
P.~Ball, \jhep 9809 (1998) 005;
P.~Ball, \jhep 9901 (1999) 010;

\bibitem{kf}
T.~Feldmann and P.~Kroll, \epjc {\bf 5} (1998) 327.

\bibitem{bf}
V.M.~Braun and I.E.~Filyanov, \zpc {\bf 48}  (1990)239.

\bibitem{sum}
 P.~Ball and V.M.~Braun, \prd {\bf 58}  (1998) 094016;
 P.~Ball and R.~Zwicky, \prd {\bf 71}   (2005) 014015.

\bibitem{charles05}
J.~Charles {\it et al.}, \epjc {\bf 41}  (2005) 1.

\bibitem{akl2}
A.~Ali, G.~Kramer and C.D.~L\"u, \prd {\bf 59}  (1999) 014005.

\bibitem{nlo}
H.-n.~Li, S.~Mishima, A.I.~Sanda, hep-ph/0508041.

\end{thebibliography}
\end{document}